\documentclass[aps,prb,twocolumn,superscriptaddress,showpacs]{revtex4}
\usepackage{color}
\usepackage{graphics}
\usepackage{subfigure}
\usepackage{longtable}
\usepackage{times}
\usepackage{graphicx}% Include figure files
\usepackage{amsmath}
\usepackage{amsbsy}
\usepackage{dcolumn}% Align table columns on decimal point
\usepackage{bm}% bold math

\newcommand{\gtwid}{\mathrel{\raise.3ex\hbox{$>$\kern-.75em\lower1ex\hbox{$\sim$}}}}
\newcommand{\ltwid}{\mathrel{\raise.3ex\hbox{$<$\kern-.75em\lower1ex\hbox{$\sim$}}}}
\begin{document}

%\title{Pairing Strength in a doped
% multi-orbital Model of the Fe-pnictides}
%

\title{Near-degeneracy of several pairing channels in multiorbital
models for the Fe-pnictides}

\author{S. Graser}
\affiliation{Department of Physics, University of Florida,
Gainesville, FL 32611, U.S.A.}

\author{T. A. Maier}
\affiliation{Center for Nanophase Materials Sciences and Computer
Science and Mathematics Division, Oak Ridge National Laboratory,
Oak Ridge, TN 37831-6494}

\author{P. J. Hirschfeld}
\affiliation{Department of Physics, University of Florida,
Gainesville, FL 32611, U.S.A.}

\author{D. J. Scalapino}
\affiliation{Department of Physics,
University of California, Santa Barbara, CA 93106-9530 USA}

\date{\today}

\begin{abstract}

%Random phase approximation (RPA) results for the coupling strength and momentum
%dependence of the gap function are discussed for electron doped tight
%binding models of the Fe-pnictides. The differences between a two-orbital and
%five-orbital tight binding fit are discussed and the relation of these results
%to present experiments and previous calculations are discussed.

Weak-coupling approaches to the pairing problem in the iron
pnictide superconductors have predicted a wide variety of
superconducting ground states.  We argue here that this is due
both to the inadequacy of certain approximations to the effective
low-energy band structure, and to the natural near-degeneracy of
different pairing channels in superconductors with many distinct
Fermi surface sheets.  In particular, we review attempts to
construct two-orbital effective band models, the argument for
their fundamental inconsistency with the symmetry of these
materials, and compare  the dynamical susceptibilities
of two and five-orbital tight-binding models.
%in two- and
%five-band  models.
We then present results for the magnetic properties, pairing
interactions, and pairing instabilities within a five-orbital tight-binding
Random Phase Approximation model. We
discuss the robustness of these results for different dopings,
interaction strengths, and variations in band structure.  Within
the parameter space explored, an anisotropic, sign-changing
$s$-wave ($A_{1g}$) state and a $d_{x^2-y^2}$ ($B_{1g}$) 
state are nearly degenerate, due to the near nesting
of Fermi surface sheets.
\end{abstract}


\maketitle

\section{Introduction}
\label{sec:1} 
\begin{figure}
\includegraphics[width=.8\columnwidth]{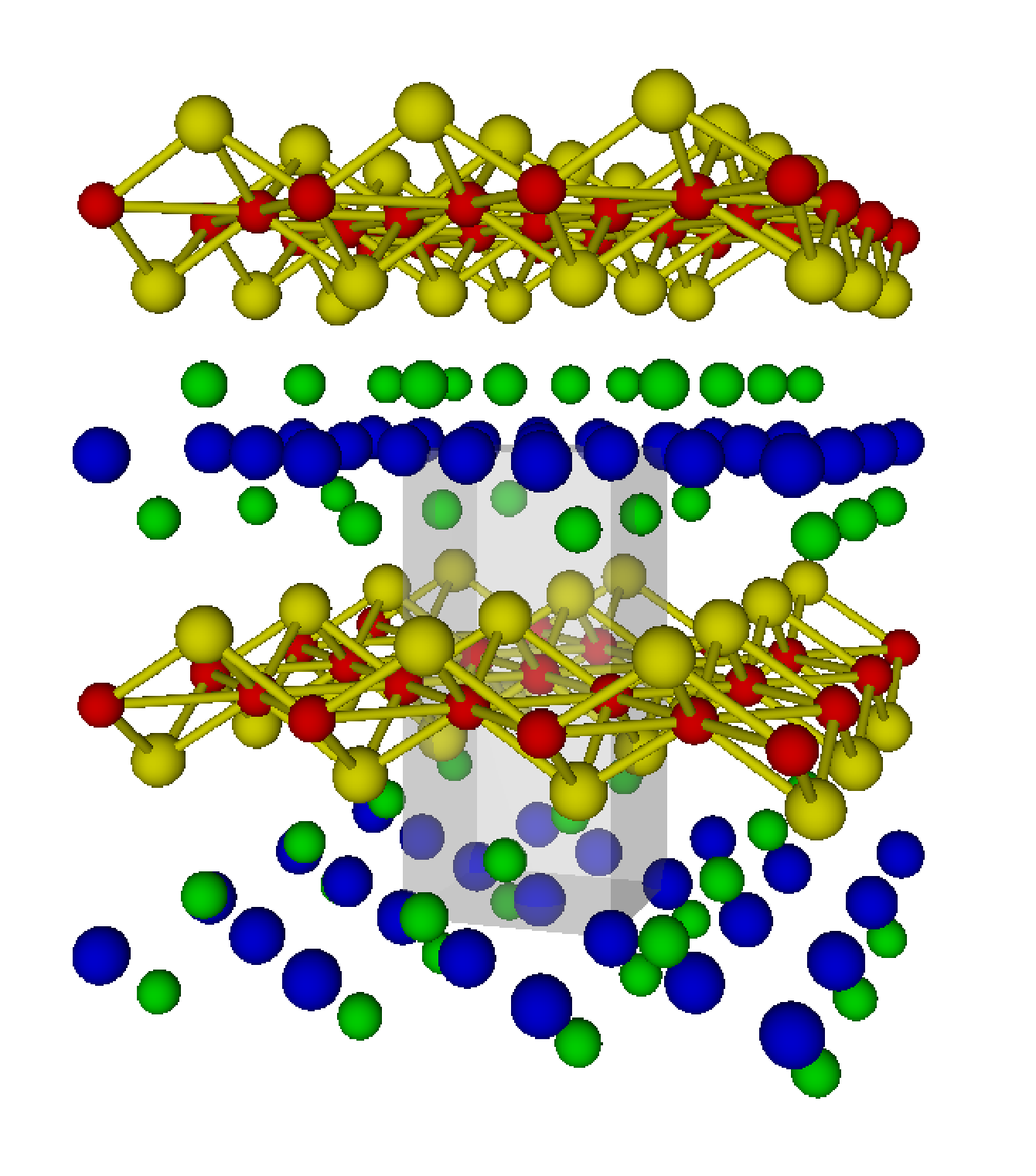}
\caption{(Color online) The crystal structure of LaOFeAs showing
the Fe-As layers with an Fe square lattice (red) and As atoms
(yellow) in a pyramidal configuration above and below the Fe
plane.} \label{fig:crystalstructure}
\end{figure}

\begin{figure}
\includegraphics[width=.6\columnwidth]{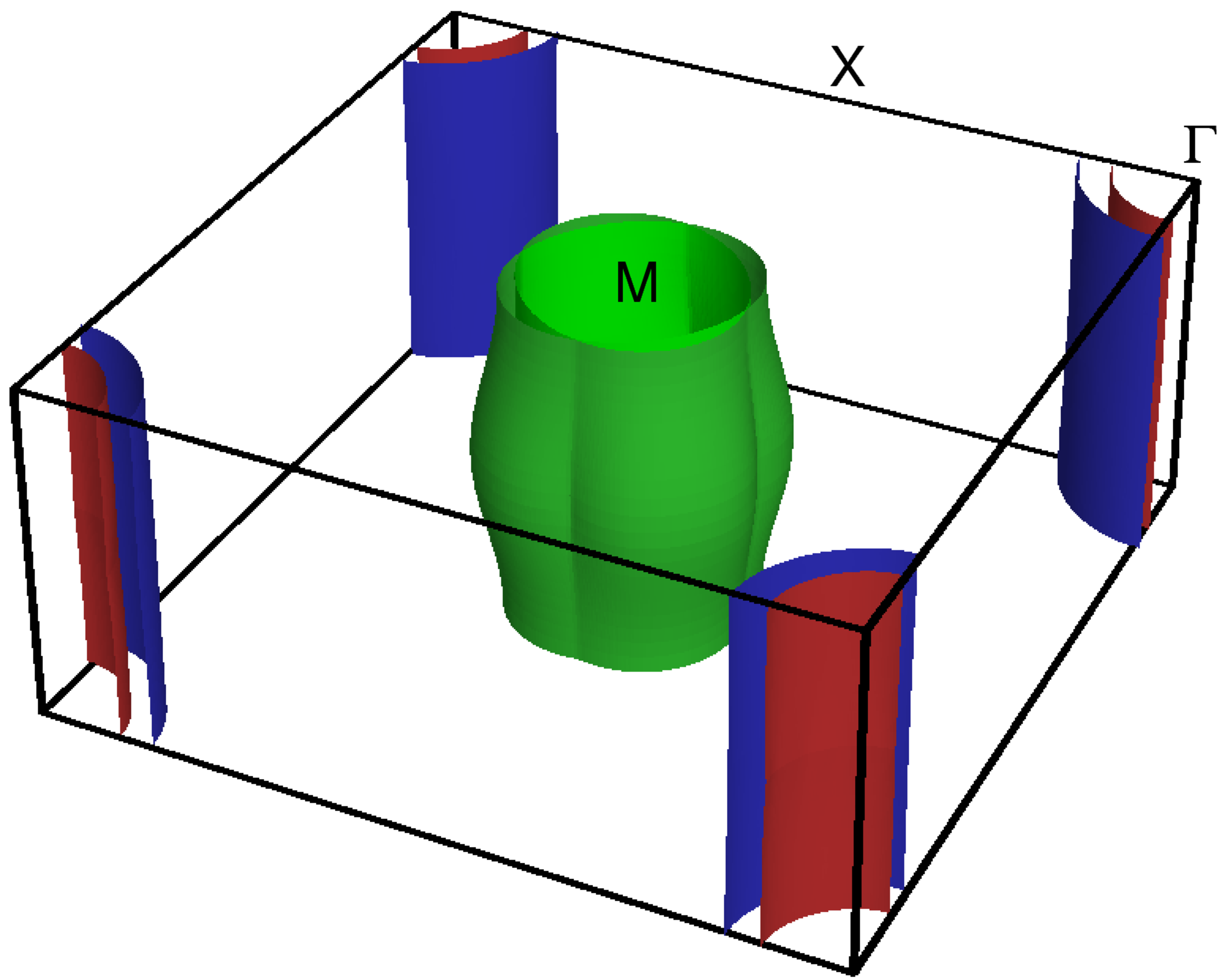}
\caption{(Color online) The Fermi surface for LaOFFeAs found in
Ref. \onlinecite{ref:Cao}.}
\label{fig:Fermisurface}
\end{figure}

The undoped Fe-pnictides are semi-metallic materials
which exhibit structural and spin density wave (SDW)
antiferromagnetic transitions. Photoemission~\cite{ref:Malaeb} and
density functional theory (DFT)
calculations~\cite{ref:Lebegue,ref:Singh,ref:Andersenpc,ref:Cao}
find that over an energy range near the Fermi energy the electron
bands are made of states from the Fe-pnictide layers
(Fig.~\ref{fig:crystalstructure}) of predominantly Fe character.
The two dimensional nature of these layers is such that the Fermi
surfaces consist of hole cylinders around the $\Gamma$ point and
electron cylinders around the $M$ point of the 2 Fe/cell Brillouin
zone (see Fig.~\ref{fig:Fermisurface}). In the undoped system, the
near nesting of the hole and electron Fermi surfaces can give rise
to a colinear antiferromagnetically ordered state within a weak
coupling approximation~\cite{ref:Dong}, and this state has been
confirmed in neutron experiments~\cite{ref:delaCruz}.  In both the
electron and hole doped cases, superconductivity appears in
proximity to or coexists with the antiferromagnetic SDW
order~\cite{ref:delaCruz,ref:Zhao,ref:Luetkens,ref:Drew}. It is
therefore natural to consider the possibility that
spin fluctuations provide the pairing mechanism and various random
phase~\cite{ref:Kuroki,ref:Qi,ref:Barzykin,ref:Bang,ref:Yanagi}
(RPA), fluctuating exchange~\cite{ref:Yao,ref:Sknepnek,ref:Ikeda}
(FLEX) and renormalization group~\cite{ref:Wang,ref:Chubukov}
calculations have been reported. These have made use of different
approximations to the band structure and obtained a variety of
different gap structures.

 At the same time, various experimental
probes of different Fe-pnictides appear to indicate quite
different symmetries of the order parameter~\cite{ref:Mu,ref:Matano,
ref:Mukuda,ref:Nakai,ref:ChenYY,ref:Kondo,
ref:Li,ref:Shan,ref:Evtushinsky,ref:Martin,
ref:Hashimoto,ref:Ding}; as discussed below,
there is a real possibility that different symmetries are realized
in different materials. It  is therefore important to investigate
the origin of the discrepancies among the various theories with
regard to  their predictions for the superconducting ground state.
In the case of the cuprates, where a single Cu-O hybridized band
predominates at the Fermi surface, different methods and band
structures all lead to $d$-wave pairing. If the situation is
qualitatively different in the Fe-pnictides, we need to isolate
which of the apparently small differences in the various
approaches is responsible for the different results obtained.  The
reward for such an effort may be considerable, since the evidence
suggests that correlation effects in these materials may be modest
compared to the cuprates, raising the possibility of a
quantitative theory of the superconducting state.

 Here we describe the results of an RPA
calculation of the pair coupling strength and the momentum
dependence of the gap function for a multiorbital tight binding
parametrization of a DFT band structure for the LaOFeAs material
by  Cao {\it et al.}~\cite{ref:Cao}.   We begin in
Sec.~\ref{sec:2} by discussing the two orbital model for the
electronic structure of the pnictides that was introduced by some
of the present authors and has been studied by various groups,
along with some of the inconsistencies associated with this
approach. In Sec.~\ref{sec:3}, we discuss our results for the
electronic structure using a five-orbital fit to DFT.   Then in
Sec.~\ref{sec:4} we examine the spin and orbital susceptibilities
and the effective pairing interaction for both models. In
Sec.~\ref{sec:5}, an effective pairing interaction strength
$\lambda$ and gap function $g(k)$ are introduced and in
Sec.~\ref{sec:6} we discuss our results for $\lambda$ and $g(k)$.
In Sec.~\ref{sec:7}, we examine the spatial and orbital structure
of the pairing interaction and the resulting pairs. Our
conclusions and comparison with previous work are contained in the
final section~\ref{sec:8}.

\section{Two-orbital model}
\label{sec:2} 
\begin{figure}
\includegraphics[width=1\columnwidth]{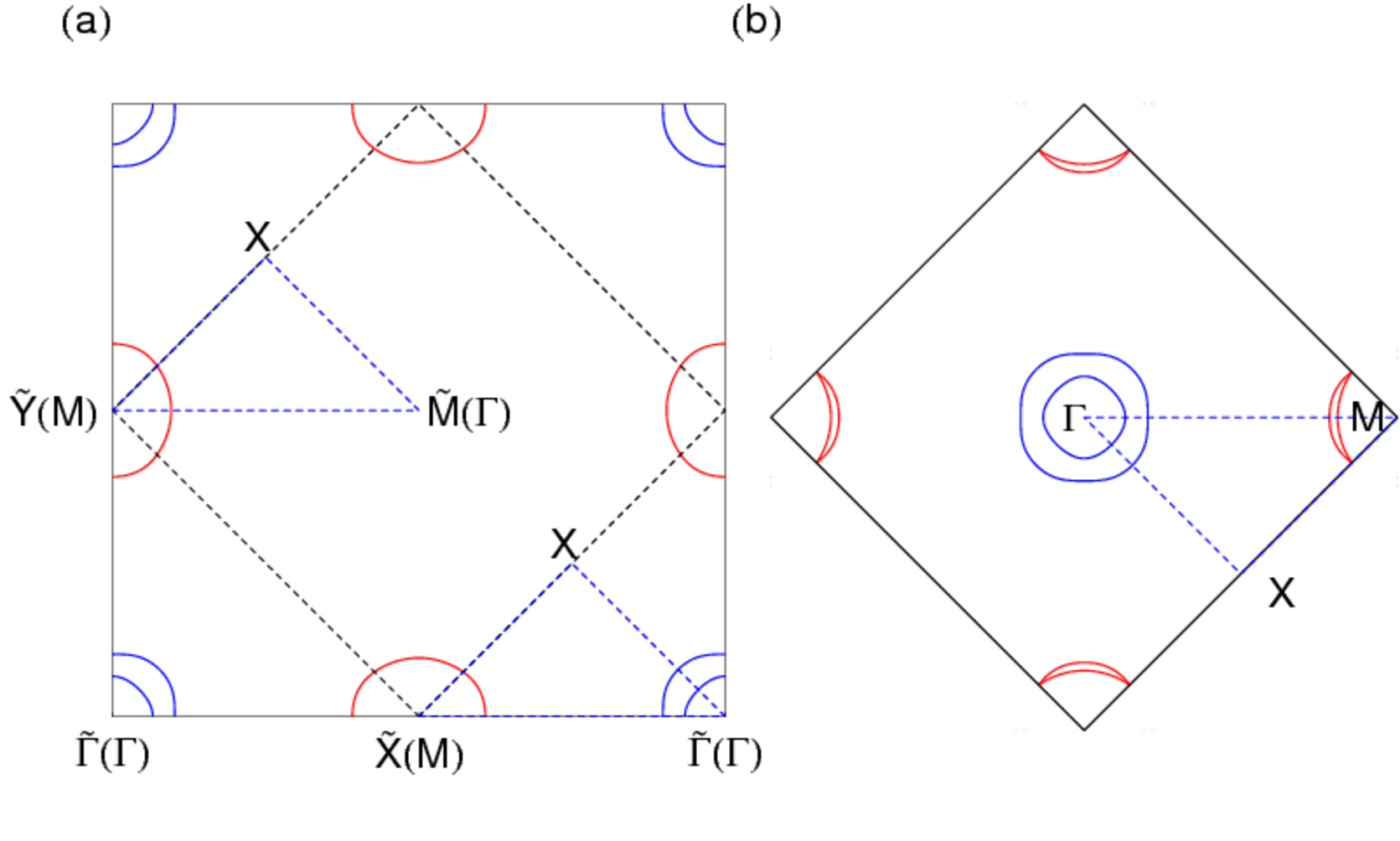}
\caption{(Color online) (a) Effective Brillouin zone for Fe-Fe
tight-binding model; (b) Correct zone. In (a) the dashed black line
shows the boundaries of the correct zone, the dashed blue
lines show the two disconnected pieces of the path
that have to be added to give a corresponding path in (b).}
\label{fig:BZ}
\end{figure}

Effective models of the band structure based
entirely on Fe orbitals should be possible because DFT tells us
that the states due to the As 4p orbitals are located
approximately 2 eV below the Fermi level.  The As orbitals allow
for hybridization with the Fe 3d states, however, and therefore an
effective Fe-Fe hopping Hamiltonian can be constructed, provided
the symmetries of the entire FeAs layer are respected. It is
conceptually simplest to work therefore with a  square lattice
with sites corresponding to Fe atoms and introduce a set of
effective hoppings between these sites.  If the primitive unit
cell is taken to be a square containing a single Fe atom, the
effective Brillouin zone is the square shown in Fig.~\ref{fig:BZ}a.
Since the true primitive unit cell contains two Fe
atoms, the true Brillouin zone is smaller by a factor of two, as
shown in Fig.~\ref{fig:BZ}b.  The Fermi surface in the correct
small zone should be obtained from the effective large zone by
folding the effective zone about the black dashed lines in
Fig.~\ref{fig:BZ}a.   Sheets around the $M$ point in Fig.~\ref{fig:BZ}a
thus correspond to sheets around the $\Gamma$ point in
Fig.~\ref{fig:BZ}b.

In Qi {\it et al.}~\cite{ref:Qi}, a 2-orbital tight binding model
was used to carry out an RPA calculation of the spin and orbital
fluctuation pairing interaction. Here we briefly describe this
model and discuss why it is insufficient to approximate the full
band structure, especially with regard to the correct orbital
weights along the Fermi surface sheets. The Hamiltonian for the
two band model can be written as
\begin{equation}
  H_0=\sum_{k\sigma}\psi^+_{k\sigma}\left[\left(\epsilon_+(k)-\mu\right)1+
    \epsilon_-(k)\tau_3+\epsilon_{xy}(k)\tau_1\right]\psi_{k\sigma}
\label{eq:1}
\end{equation}
Here, $\tau_i$ are Pauli matrices and $\psi^+_{k\sigma}=\left(d^+_{x\sigma}(k),
d^+_{y\sigma}(k)\right)$ is a two-component field which describes the two
``$d_{xz}$" and ``$d_{yz}$" orbitals. The energies $\epsilon_\pm(k)$ and
$\epsilon_{xy}(k)$ are parameterized in terms of four hopping parameters $t_i$:
\begin{eqnarray}
  \epsilon_+(k) &=& -(t_1+t_2)(\cos k_x+\cos k_y)-4t_3\cos k_x\cos k_y\nonumber\\
    \epsilon_-(k) &=& -(t_1-t_2)(\cos k_x-\cos k_y)\\
    \epsilon_{xy}(k) &=& -4t_4\sin k_x\sin k_y\nonumber
\label{eq:2}
\end{eqnarray}
Diagonalizing $H_0$ one has
\begin{equation}
  \psi_{s\sigma}(k)=\sum_{\nu=\pm}a^s_\nu(k)\gamma_{\nu\sigma}(k)
\label{eq:3}
\end{equation}
Here, $\gamma_{\nu\sigma}(k)$ destroys an electron in the $\nu=\pm$ band with
spin $\sigma$ and
\begin{eqnarray}
  a^x_-(k) &=& \left[\frac{1}{2}\left(1+
    \frac{\epsilon_-(k)}{\sqrt{\epsilon^2_-(k)+\epsilon^2_{xy}(k)}}\right)\right]^{1\over 2}
    =a^y_+(k)\nonumber\\
    a^x_+(k) &=& \left[\frac{1}{2}\left(1-
    \frac{\epsilon_-(k)}{\sqrt{\epsilon^2_-(k)+\epsilon^2_{xy}(k)}}\right)\right]^{1\over 2}
    =-a^y(\kappa)\nonumber\\&&
\label{eq:4}
\end{eqnarray}

\begin{figure}
\includegraphics[width=.6\columnwidth]{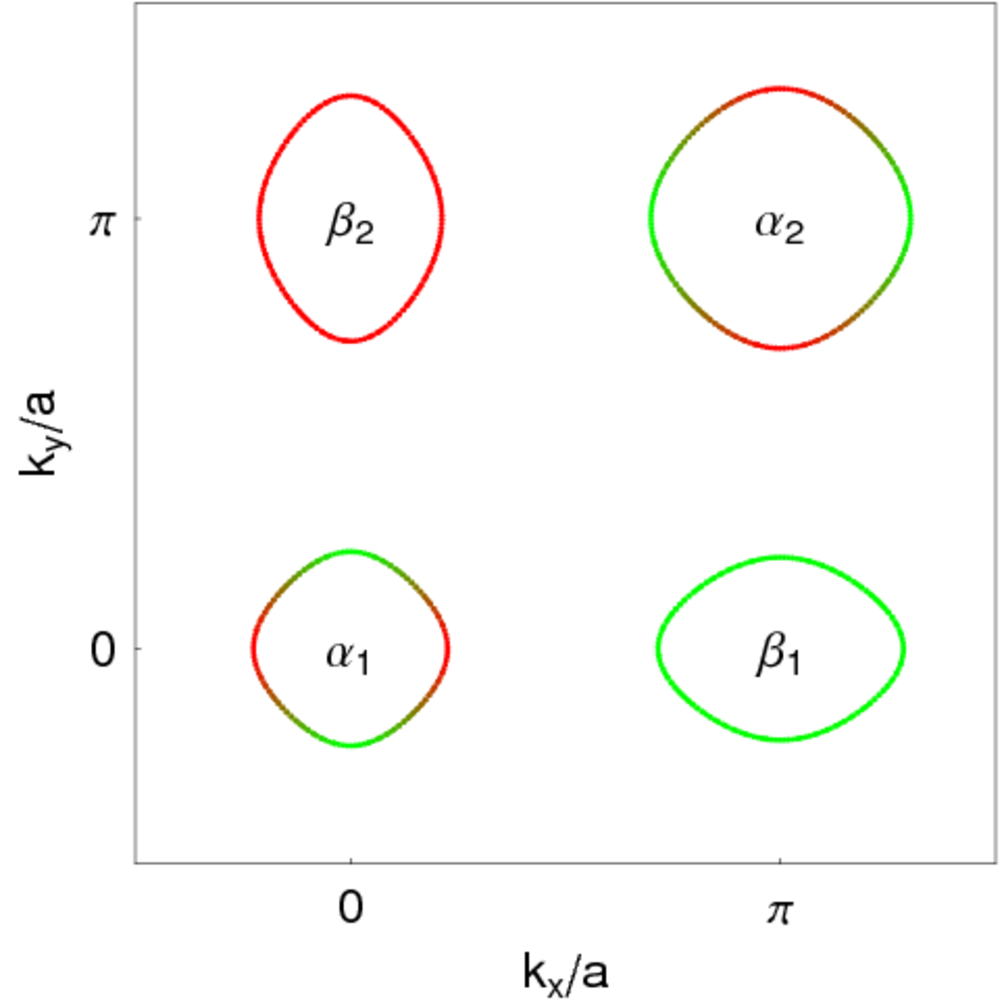}
\caption{(Color online) Fermi surface for the 2-orbital
$d_{xz}$-$d_{yz}$ model with $\mu=1.5$ showing the $\alpha_2$ FS
sheet around $k=(\pi,\pi)$ rather than around $k=(0,0)$. The
orbital contributions of the band states that lie on the different
FS sheets are shown color coded: $d_{xz}$ (red) and $d_{yz}$
(green).} \label{fig:FS2band}
\end{figure}

With the addition of a chemical potential the Hamiltonian takes the following form
\begin{equation}
  H_0=\sum_{k\sigma} \sum_{\nu=\pm}\left(E_\nu(k)-\mu\right)\gamma^+_{\nu\sigma}(k)\gamma_{\nu\sigma}(k)
\label{eq:5}
\end{equation}
where the band energies are
$E_\pm(k)=\epsilon_+(k)\pm\sqrt{\epsilon^2_-(k)+
\epsilon^2_{xy}(k)}$. Measuring energy in units of $|t_1|$, for
$t_1=-1$, $t_2=1.3$, $t_3=t_4=-0.85$ and $\mu=1.5$ we exhibit the Fermi
surface  in Fig.~\ref{fig:FS2band}, displayed in the large
effective zone.  To obtain the corresponding Fermi surface in the
small zone, a folding across the line joining $(\pi,0)$ and $(0,\pi)$
is required, as discussed above.

The Fermi surfaces $\alpha_1$ around $(0,0)$ and $\alpha_2$ around
$(\pi,\pi)$ are hole pockets associated with $E_-(k_f)=0$ and the
$\beta_1$ and $\beta_2$ Fermi surfaces around $(\pi,0)$ and
$(0,\pi)$ are electron pockets from $E_+(k_f)=0$.  We note that
the displacement of the $\alpha_2$ Fermi surface from the
$\Gamma$-point to $(\pi,\pi)$ is an artifact of the 2-orbital
approximation.  As shown by various DFT
calculations~\cite{ref:Singh,ref:Cao,ref:Andersenpc} and noted by
Lee and Wen~\cite{ref:LeeWen}, the orbital states that
have significant weight  near the $\beta$ Fermi surfaces include, in
addition to the $d_{xz}$ and $d_{yz}$ orbitals, a $d_{xy}$
orbital. In addition, while the Fermi surface sheets shown in
Fig.~\ref{fig:FS2band} fold down to give two hole Fermi surfaces
around the $\Gamma$ point of the 2Fe/cell Brillouin zone, there
should in fact be two hole Fermi surfaces around the $\Gamma$
point of the large, effective Brillouin zone. This is known from
the wave functions found in the band structure calculations.
Finally, the 2-orbital model lacks the flexibility to fit the
Fermi velocities found in the band structure calculations, giving
Fermi velocities on the electron $\beta$ Fermi surfaces that are
anomalously small compared to the velocities on the hole Fermi
surfaces.

\section{5-orbital Model}
\label{sec:3}

 In principle, one could capture the correct behavior
near the $\alpha$ and $\beta$ Fermi surface sheets by treating a
3-orbital $d_{xz}$, $d_{yz}$, $d_{xy}$ model. However, with short
range hoppings, this leads to the appearance of an extra
unphysical Fermi surface and a fourth orbital is required to
remove it~\cite{ref:LeeWen}. In addition, recent theoretical
calculations using 5 Wannier d-orbits per Fe site find that one
can obtain an excellent representation of the electronic structure
within a $\pm$2eV window of the Fermi energy~\cite{ref:Kuroki}.
Furthermore, the values of the Coulomb interaction parameters
obtained for the Wannier basis are such that the average Coulomb
interaction is small compared to the bandwidth, implying that one
is dealing with a weakly coupled system~\cite{ref:anisimov}.

 At this point it therefore seems best to use
all five $d$ orbitals in developing a tight-binding
model\cite{ref:Kuroki}. Here, using a Slater-Koster based
parametrization which respects the symmetry of the FeAs layers, we
fit a five-orbital ($d_{xz}$, $d_{yz}$, $d_{xy}$, $d_{x^2-y^2}$, and
$d_{3z^2-r^2}$) tight binding model to the DFT band structure
determined by Cao {\it et al.}~\cite{ref:Cao}. We will use an
orbital basis that is aligned parallel to the nearest neighbor
Fe-Fe direction rather than the Fe-As direction. With this choice
we avoid the necessity of introducing a second, rotated coordinate
system in addition to the one that is used to describe the single
Fe unit cell.

The Hamiltonian for the 5 band model takes the following form
\begin{equation}
H_0 = \sum_{k\sigma} \sum_{mn} \left( \xi_{mn} (k) + \epsilon_m \delta_{mn} \right)
d_{m\sigma}^\dagger(k) d_{n\sigma}(k)
\end{equation}
Here $d_{m,\sigma}^\dagger(k)$ creates a particle with momentum
$k$ and spin $\sigma$ in the orbital $m$. The kinetic energy terms
$\xi_{mn} (k)$ together with the parameters for a 5 band tight
binding fit of the DFT band structure by  Cao {\it et al.} are
listed in the appendix. A diagonalization of this Hamiltonian
yields the eigenenergies and the matrix elements analogously to
the two band case discussed above.

\begin{figure}
\includegraphics[width=.9\columnwidth]{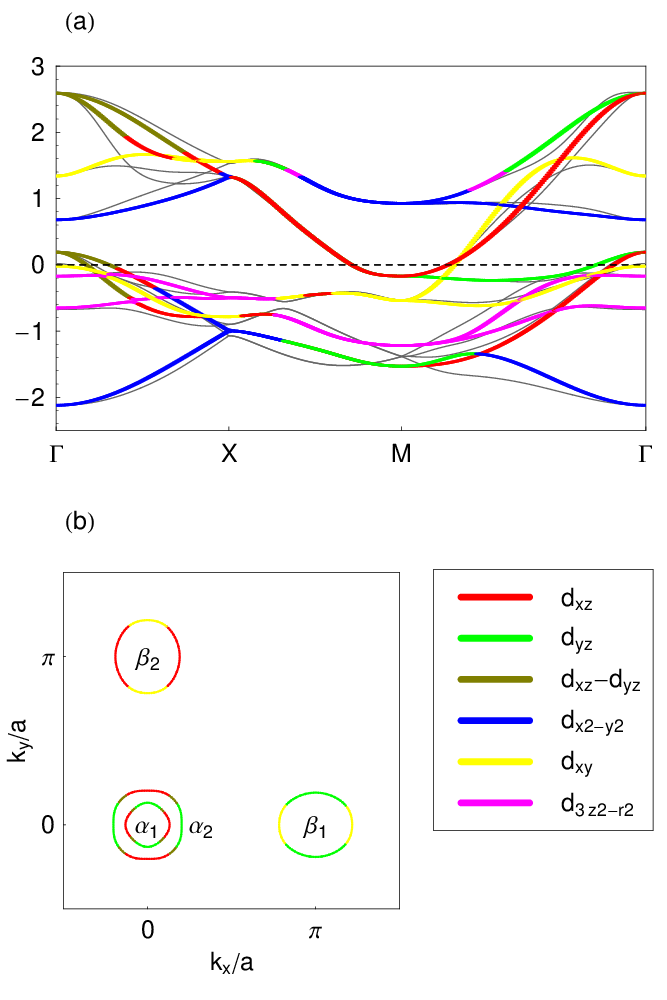}
\caption{(Color online) (a) The backfolded band structure for the
5 band model with $\Gamma$, $X$, and $M$ denoting the symmetry
points in the real Brillouin zone corresponding to the 2 Fe unit
cell. The main orbital contributions are shown by the
following colors: $d_{xz}$ (red), $d_{yz}$ (green), $d_{xy}$
(yellow), $d_{x^2-y^2}$ (blue), $d_{3z^2-r^2}$ (magenta), and a
strongly hybridized $d_{xz}$-$d_{yz}$ band (brown). The gray lines
show the correct DFT band structure calculated by Cao {\it et al.}
(b) The FS sheets of the 5 band model for the undoped compound
($x=0$).} \label{fig:BSFS5band}
\end{figure}

\begin{figure}
\includegraphics[width=1.\columnwidth]{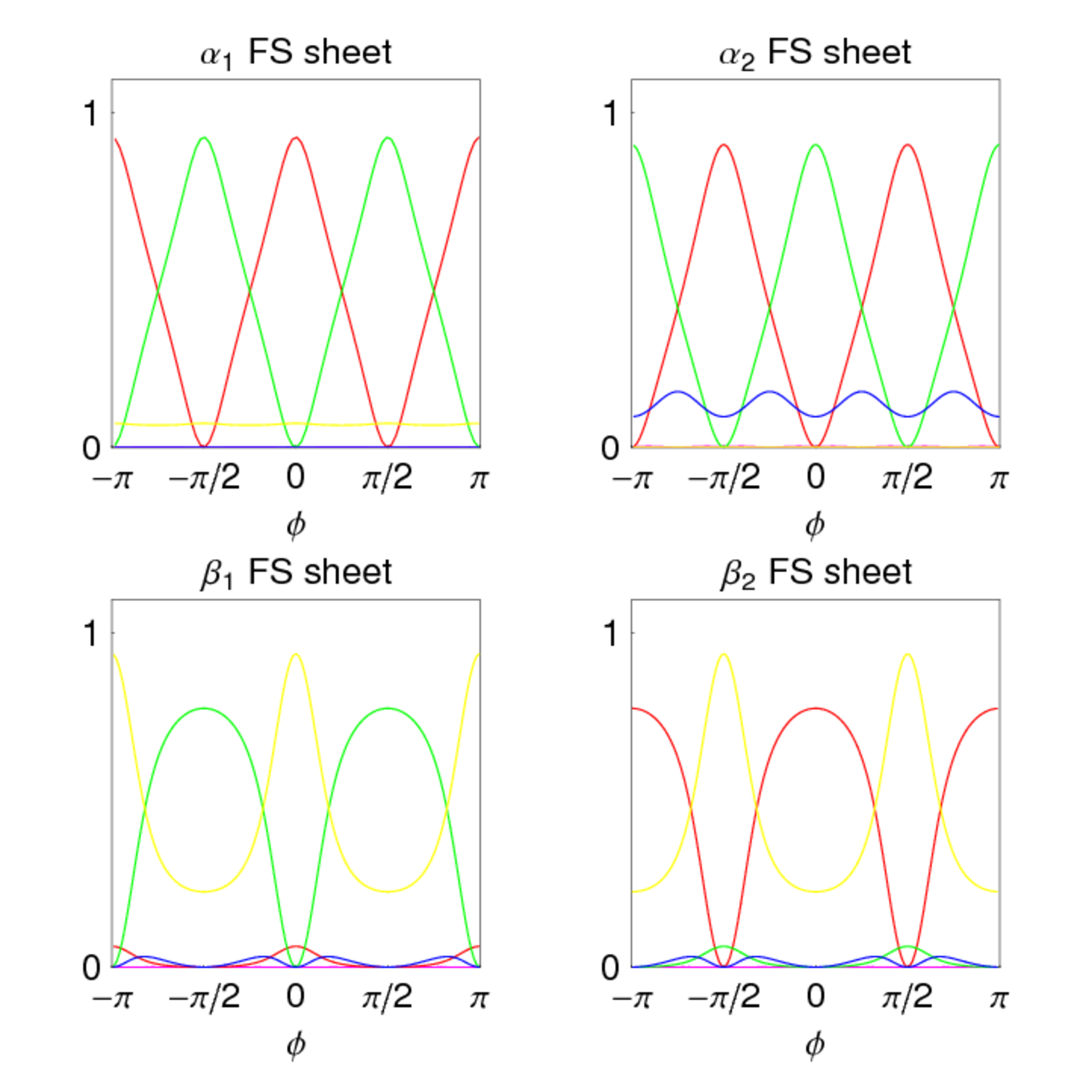}
\caption{(Color online) The orbital weights as a function of the
winding angle $\phi$ on the different Fermi surface sheets. The
different colors refer to $d_{xz}$ (red), $d_{yz}$ (green),
$d_{xy}$ (yellow), $d_{x^2-y^2}$ (blue), $d_{3z^2-r^2}$
(magenta).} \label{fig:OrbitalW5band}
\end{figure}

In Fig.~\ref{fig:BSFS5band} (a), we have plotted the resulting band
structure in the backfolded ``small" Brillouin zone while the
Fermi surface sheets for zero doping are shown in
Fig.~\ref{fig:BSFS5band} (b). The colors correspond to the dominant
orbital weight of each band in momentum space. The gray lines
represent the DFT band structure  by  Cao {\it et al.} and the
comparison shows, that the 5 band fit approximately reproduces the
DFT bands, especially in the vicinity of the Fermi level. It is
obvious that the $d_{xy}$ contribution plays an important role in
building up the electron-like Fermi surfaces ($\beta$ sheets)
around the $M$ point of the ``small" Brillouin zone. In the
unfolded Brillouin zone (Fig.~\ref{fig:BSFS5band} (b) the $d_{xy}$
orbital and $d_{yz} (d_{xz})$ orbital contribute the dominant
weights to the band states at the $\beta_1 (\beta_2)$ Fermi
surface. To confirm this we also show the orbital weights as a
function of the winding angle on the different Fermi surfaces in
Fig.~\ref{fig:OrbitalW5band}.

\section{The spin and charge susceptibilities}
\label{sec:4}
\subsection{Noninteracting susceptibilities}

In this section we examine the RPA enhanced
spin and charge susceptibilities for the multiorbital model that we
introduced in the previous section. The spin operator for an orbital $s$ is defined as
\begin{equation}
\vec{S}_s (q) = \frac{1}{2} \sum_{k,\alpha\beta}
d_{s\alpha}^\dagger (k+q) \vec{\sigma}_{\alpha\beta}
d_{s\beta} (k)
\end{equation}
where $\alpha$ and $\beta$ are spin indices. The spin susceptibility
can then be calculated from  the Matsubara spin-spin correlation function
 \begin{equation}
(\chi_1)_t^s(q,i\omega)  = \frac{1}{3} \int_0^\beta d\tau\; e^{i\omega \tau}\left\langle
T_\tau \vec{S}_s (q,\tau) \vec{S}_t (-q,0) \right\rangle
\end{equation}
with $\tau$ the imaginary time and $\omega$ a Matsubara frequency.
In the same way we can define the Fourier component of the charge
density for the orbital $s$ as
\begin{equation}
n_s (q) = \sum_{k,\alpha\beta} d_{s\alpha}^\dagger (k+q)
d_{s\beta} (k) \delta_{\alpha\beta}
\end{equation}
and we can calculate the charge susceptibility from
\begin{equation}
(\chi_0)_t^s (q,i\omega)  =  \int_0^\beta d\tau\; e^{i\omega \tau}\left\langle
T_\tau n_s (q,\tau) n_t (-q,0) \right\rangle
\end{equation}
In a more general formulation the susceptibilities are functions of four
orbital indices. For the non-interacting case $(\chi_0)_{st}^{pq}$ and
$(\chi_1)_{st}^{pq}$ are equivalent and can be written
\begin{eqnarray}
\chi_{st}^{pq} (q,i\omega) & = &  \int_0^\beta d\tau\; e^{i\omega \tau}
\sum_{kk'} \sum_{\alpha\beta} \langle
T_\tau d_{p\alpha}^\dagger (k,\tau)    \\
& & d_{q\alpha} (k+q,\tau) d_{s\beta}^\dagger (k',0) d_{t\beta} (k'-q,0)
\rangle  \nonumber
\end{eqnarray}
Now we can derive an explicit expression for the non-interacting susceptibilities from
\begin{equation}
\chi_{st}^{pq} (q,i\omega) = - \frac{1}{N\beta} \sum_{k,i\omega_n} G_{sp} (k,i\omega_n)
G_{qt} (k+q,i\omega_n+i\omega)
\end{equation}
where $N$ is the number of Fe lattice sites, $\beta=1/T$, and the
spectral representation of the Green's function is given as
\begin{equation}
G_{sp} (k,i\omega_n) = \sum_\mu \frac{a_\mu^s(k) a_\mu^{p*}(k)}{i \omega_n - E_\mu(k)}
\end{equation}
Here the matrix elements $a_\mu^s(k) = \langle s | \mu k \rangle$
connect the orbital and the band space and are the
components of the eigenvectors resulting from the diagonalization
of the initial Hamiltonian. Performing the Matsubara frequency
summation and setting $i \omega_n \rightarrow \omega + i 0^+$ we
find the retarded susceptibility
\begin{eqnarray}
\chi_{st}^{pq} (q,\omega) & = & - \frac{1}{N} \sum_{k,\mu\nu} \frac{a_\mu^s(k) a_\mu^{p*}(k) a_\nu^q(k+q) a_\nu^{t*}(k+q)}
{ \omega + E_\nu(k+q) - E_\mu(k) + i 0^+} \nonumber \\
& & \times \left[ f(E_\nu(k+q)) - f(E_\mu(k)) \right]
\end{eqnarray}

\subsection{Random phase approximation for multiorbital system}

\begin{figure}
\includegraphics[width=1.\columnwidth]{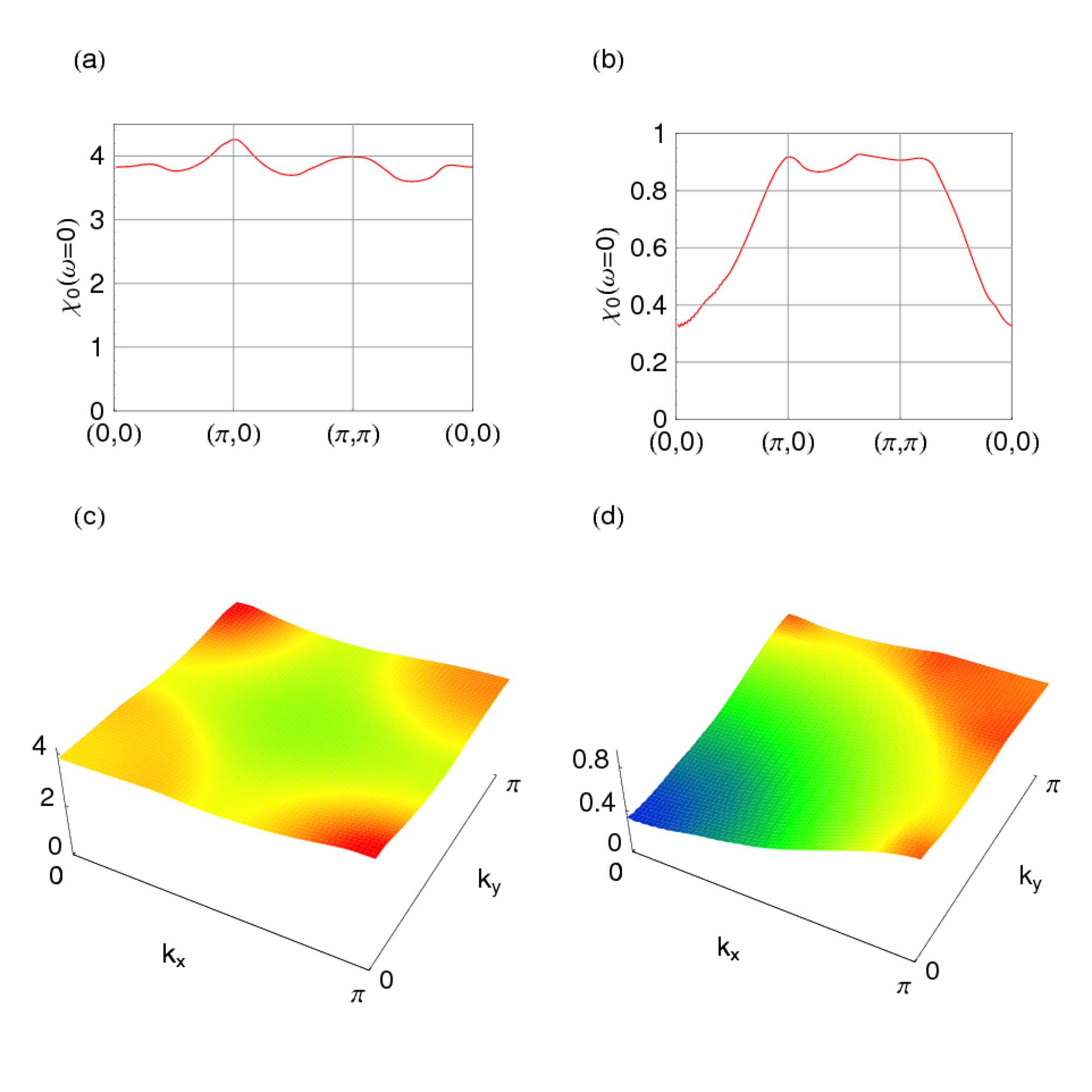}
\caption{(Color online) The bare spin susceptibility $\chi_S(q)$ calculated for the
undoped 5 orbital model without (a,c) and with (b,d) the matrix elements.
While (a) and (b) show cuts of the susceptibility
along the main symmetry directions, (c) and (d) show the
susceptibility in the first quadrant of the effective Brillouin zone.}
\label{fig:Chi0}
\end{figure}

Now we  consider Coulomb interactions of the electrons on the same
Fe atom in an RPA framework. We  distinguish between an
intraorbital interaction $U$ of electrons in the same orbital and
an interorbital interaction of electrons in different orbitals
$V$. We can also take the Hund's rule coupling into account that
favors the parallel alignment of electron spins on the same ion
and is described by an energy $J>0$ and the pair hopping energy
denoted by $J'$.  These interactions are generated automatically
in multiorbital models with general two-body interactions
using a Hubbard-type approach restricted to intrasite
processes~\cite{ref:oles,ref:Takimoto,ref:kubo}.  Thus in general one  can write
\begin{eqnarray}
H_{int} & = & U \sum_{is} n_{i,s\uparrow} n_{is\downarrow} + \frac{V}{2}
\sum_{i,s,t\neq s} n_{is} n_{it} -  \frac{J}{2}  \sum_{i,s,t\neq s}
\vec{S}_{is}\cdot \vec{S}_{it} \nonumber \\
& & +  \frac{J'}{2}  \sum_{i,s,t\neq s}  \sum_\sigma c_{is\sigma}^\dagger
c_{is\bar{\sigma}}^\dagger c_{it\bar{\sigma}} c_{it\sigma}
\end{eqnarray}
where $n_{is} =  n_{i,s\uparrow} + n_{is\downarrow}$.  We have
separated the intraorbital exchange $J$ and ``pair hopping" term
$J'$ for generality, but note that if they are generated from a
single two-body term they are related by $J'=J/2$.   If we now
split the interaction Hamiltonian into singlet ($H_{int}^{(s)}$),
triplet ($H_{int}^{(t)}$) and pair ($H_{int}^{(p)}$) channels, we
find
\begin{equation}
H_{int}^{(s)} = \sum_{i,st} \left[ \frac{U}{4} \delta_{st} +  \frac{V}{2}(1- \delta_{st})\right] n_{is} n_{it}
\end{equation}
and
\begin{equation}
H_{int}^{(t)} = - \sum_{i,st} \left[ \frac{U}{12} \delta_{st} +  \frac{J}{8}(1- \delta_{st})\right] \vec{\sigma}_{is}
\vec{\sigma}_{it}
\end{equation}
where $\vec{\sigma}_{is} = 2 \vec{S}_{is}$, as well as
\begin{equation}
H_{int}^{(p)} = \sum_{i,st} \sum_\sigma \frac{J'}{2} (1-\delta_{st})  c_{is\sigma}^\dagger
c_{is\bar{\sigma}}^\dagger c_{it\bar{\sigma}} c_{it\sigma}
\end{equation}
The RPA susceptibilities are obtained in the form of Dyson-type
equations as
\begin{equation}
(\chi_{0}^{RPA})_{st}^{pq} = \chi_{st}^{pq} - (\chi_{0}^{RPA})_{uv}^{pq} (U^c)_{wz}^{uv} \chi_{st}^{wz}
\end{equation}
and
\begin{equation}
(\chi_{1}^{RPA})_{st}^{pq} = \chi_{st}^{pq} +
(\chi_{1}^{RPA})_{uv}^{pq} (U^s)_{wz}^{uv} \chi_{st}^{wz},
\end{equation}
where repeated indices are summed over. Here the non-zero
components of the matrices $U^c$ and $U^s$ are given as
\begin{equation}
(U^c)_{aa}^{aa} = U, \; (U^c)_{bb}^{aa} = 2V, \; (U^c)_{ab}^{ab} = \frac{3}{4} J - V, \; (U^c)_{ab}^{ba} = J' \nonumber
\end{equation}
and
\begin{equation}
(U^s)_{aa}^{aa} = U, \; (U^s)_{bb}^{aa} = \frac{1}{2}J, \; (U^s)_{ab}^{ab} = \frac{1}{4} J + V, \; (U^s)_{ab}^{ba} = J' \nonumber
\end{equation}
where $a\neq b$.  Our notation of the interaction parameters $U$,
$V$, $J$ and $J'$ can be compared to the notation in Kubo
~\cite{ref:kubo} as
\begin{equation}
\tilde{U}=U, \; \tilde{U}'=V+\frac{J}{4}, \; \tilde{J} = \frac{J}{2}, \tilde{J}'=J'
\end{equation}
 where $\tilde{U}$, $\tilde{V}$, $\tilde{J}$ and $\tilde{J}'$
denote the interaction parameters introduced in
Ref.~\onlinecite{ref:kubo}. In Fig.~\ref{fig:Chi0} we show results
for the static, homogeneous bare spin susceptibility
\begin{equation}
\chi_S({\bf q})=\frac{1}{2} \sum_{sp} (\chi_1)_{ss}^{pp}({\bf
q},0)\label{eq:spinsusc}
\end{equation}
as a function of ${\bf q}$ in the first quadrant of the effective
``large" Brillouin zone as well as cuts along the main symmetry
directions. We compare calculations that have been performed
correctly including the matrix elements to calculations where the
matrix elements have been considered to be constant $a_\mu^s(k)
a_\mu^{p*}(k)=1/5$, as has been reported in certain {\it ab
initio} electronic structure calculations. Here we used the five
band fit discussed in the previous section and chose a chemical
potential that corresponds to the undoped compound ($x=0$).
As can be seen from Fig.~\ref{fig:Chi0} the matrix elements play a very important
role by ``filtering" special structures of $\chi_S(q)$ in momentum space, e.g. suppressing
the weight for small $q$. The matrix elements for the $d_{x^2-y^2}$ and $d_{3z^2-r^2}$
are very small on the Fermi surfaces (Fig.~\ref{fig:OrbitalW5band}) so that the actual
$\chi_S(q)$ is reduced when these matrix elements are properly taken into account.
Neglecting the matrix elements will lead to a wrong
result and generally to a higher value and more homogeneous distribution of the
susceptibility in momentum space. Important features like the $Q=(\pi,0)$--peak,
that is responsible for the antiferromagnetic SDW instability, can be under- or overestimated.
\begin{figure}
\includegraphics[width=1.\columnwidth]{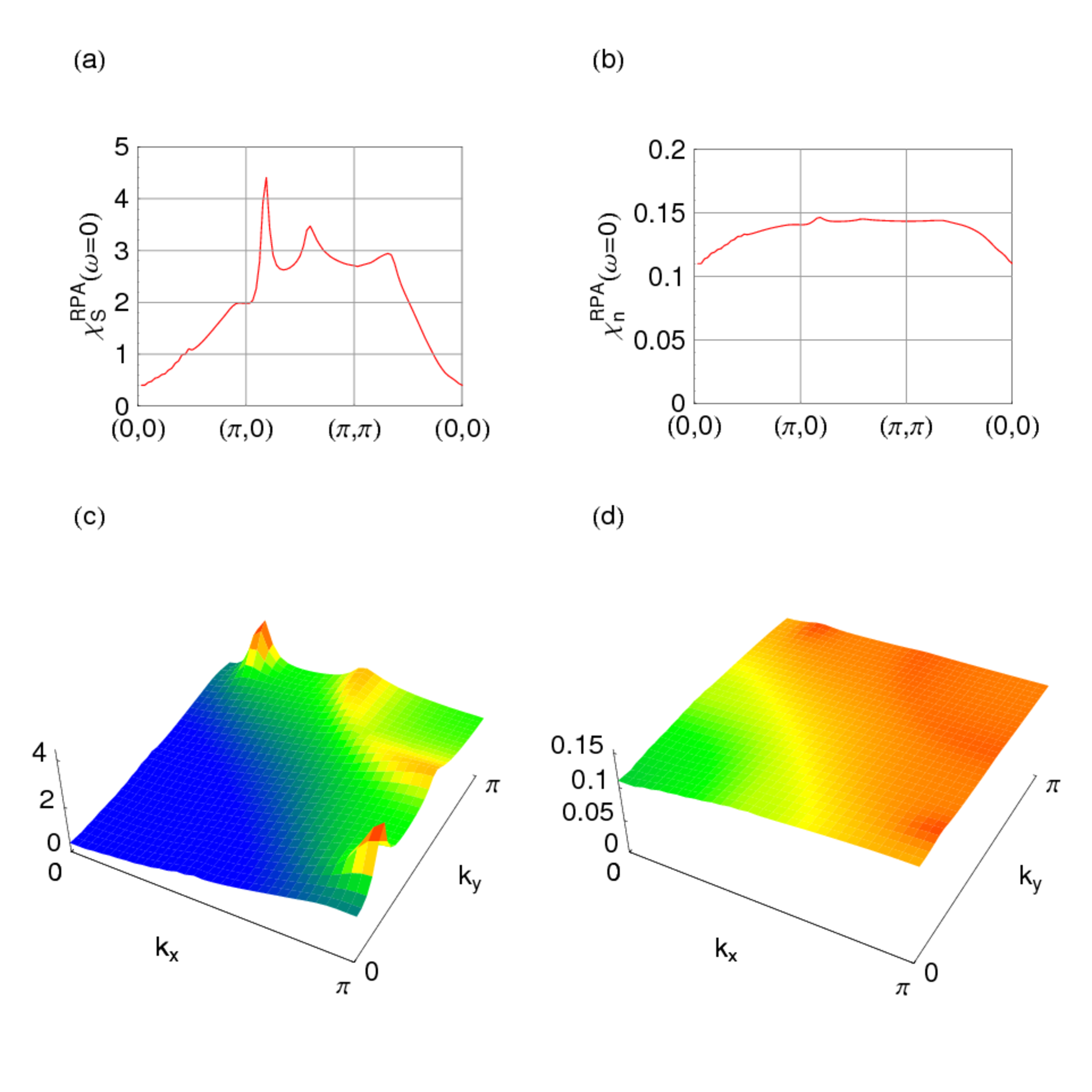}
\caption{(Color online) The RPA enhanced susceptibilities calculated for the
electron-doped compound ($x=0.125$). The interaction parameters have been chosen
as $U=V=1.65$ and $J=0$. (a) and (c) are plots of the spin susceptibility,
(b) and (d) are plots of the charge susceptibility.}
\label{fig:ChiSChin}
\end{figure}

For the single band susceptibility the inclusion of interactions
within an RPA approach is known to enhance existing structures in
the bare susceptibility as $U \chi_0'(q)$  approaches 1. In the
case of a multiorbital susceptibility with onsite intraorbital
repulsion $U$, onsite interorbital repulsion $V$ and Hund's rule
coupling $J$ it is not obvious how the different structures in the
spin and in the charge susceptibility are changed by the variation
of these three parameters. For a simplified and more transparent
discussion we will first study the susceptibility as a function of
$U$ and we will choose $V=U$ and $J=J'=0$. In
Fig.~\ref{fig:ChiSChin} we compare the spin and charge
susceptibilities of the electron-doped compound with $x=0.125$ for a
value of $U=1.65$ that is chosen such that the main peaks in the
spin susceptibility are considerably enhanced. Here, as
elsewhere in this paper, energy units are in eV.  For this value
of $U$ we find that the charge susceptibility is more than one
order of magnitude smaller than the spin susceptibility. In
addition the charge susceptibility has no pronounced structures in
momentum space whereas the spin susceptibility shows two distinct
peaks between $(\pi,0)$ and $(\pi,\pi)$. While in the undoped
compound (see Fig.~\ref{fig:Chi0}) we find the main peak in the
spin susceptibility at the antiferromagnetic wave vector
$Q=(\pi,0)$ (corresponding to $(\pi,\pi)$ in the ``small" BZ) we
observe a shift of this peak towards an incommensurate wave vector
$Q^*=(\pi,0.16\pi)$ in the electron-doped compound ($x=0.125$).
This shift can be attributed to an imperfect nesting of the
$\alpha$ and $\beta$ FS sheets due to the opposite growth of the
electron and hole sheets with doping.
\begin{figure}
\includegraphics[width=.8\columnwidth]{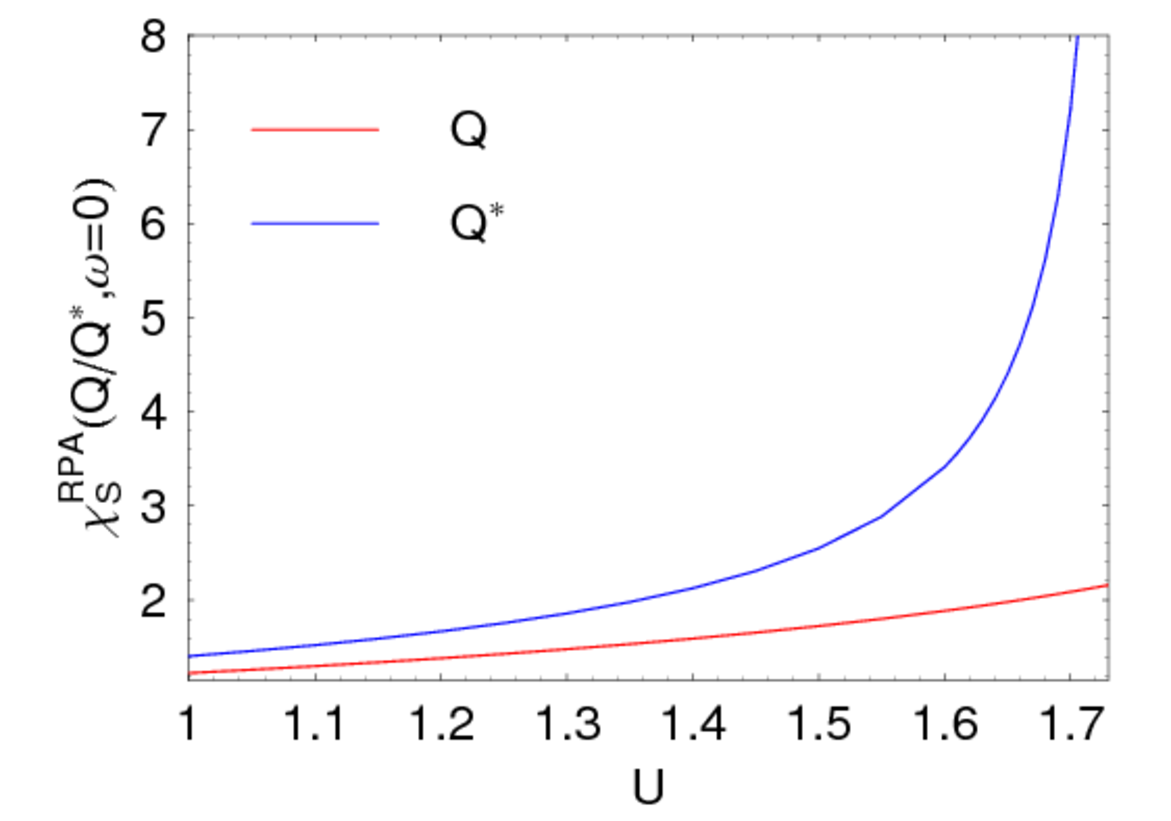}
\caption{(Color online) The RPA enhanced spin susceptibility
$\chi_S(Q,\omega=0)$ as a function of the interaction strength
$U=V$ for the electron-doped compound ($x=0.125$) at two different
momenta $Q=(\pi,0)$ and $Q^*=(\pi,0.16\pi)$.}
\label{fig:ChiRPAvsU}
\end{figure}

When we finally compare in Fig.~\ref{fig:ChiRPAvsU} the change of the spin susceptibility
as a function of $U$ for the antiferromagnetic wave vector $Q=(\pi,0)$ to the one for the
incommensurate wave vector $Q^*=(\pi,0.16\pi)$
we find a moderate and nearly linear increase of $\chi_S(Q)$ while $\chi_S(Q^*)$ diverges if we approach
the critical value of $U$.

\vskip .2cm \subsection{$T$ dependence of $\chi$ and
$(T_1T)^{-1}$}

The temperature dependence of the magnetic susceptibility and NMR
spin-lattice relaxation rate in the paramagnetic phase  are
immediate tests of our RPA calculation of $\chi_S({\bf
q},\omega)$.  The susceptibility of the F-doped LaOFeAs 1111
material which we study here has been measured both by direct
magnetization and NMR
methods~\cite{ref:RKlingeler,ref:Grafe,ref:Ahilan} and found to
increase quasi-linearly up to several hundred degrees K.  In
addition, Zhang {\it et al.} have pointed out that this behavior
appears to hold for the 122 class of materials as well, and that
the slope of the linear-$T$ behavior appears to be roughly
doping-independent~\cite{ref:DHLeelinearchi}.  Within our theory,
we can calculate the homogeneous, static bare spin susceptibility
%in units of $\mu_B^2$
by summing over spins and using
Eq.~\ref{eq:spinsusc}.  At $T=0$ it reduces to
\begin{equation}
\chi_0\equiv 2 \chi_S = 2\sum_\nu N_\nu(0),
\end{equation}
which is of course just the Pauli susceptibility proportional to
the total density of states at the Fermi level ($N_\nu(0)$ is the
single-spin density of states at the Fermi level in band $\nu$.).
At finite temperatures one might assume that the susceptibility of
an itinerant electron system should decrease.  This is the usual
case for a single parabolic band, but in the presence of band
structure effects the susceptibility may first increase.  For
example, in a single band system the band structure enters in a
simple way as $\chi_0(T)\sim \chi_0(0) +
[(N''(0)N(0)-N'(0)^2)/N(0)^2]T^2$, where $N'$, $N''$, are the
derivatives of the density of states at the Fermi level. Our
situation is also somewhat unusual as the chemical potential sits
in a region of rapidly varying density of states, so one may
expect the $T$ dependence of the susceptibility to differ from the
textbook results.

\begin{figure}
\includegraphics[width=1\columnwidth]{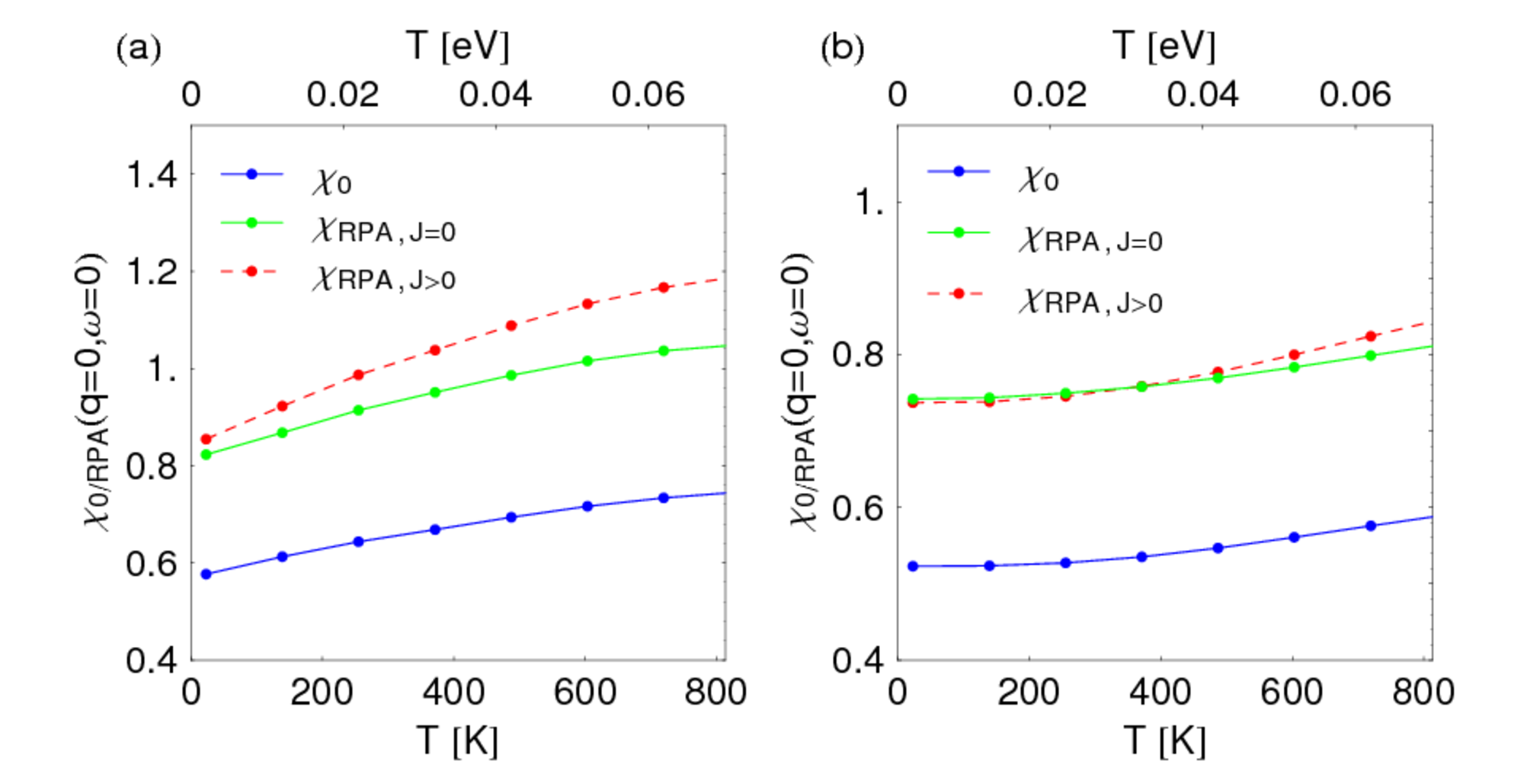}
\caption{(Color online) The  susceptibility $\chi_0(T) =
2\chi_S({\bf q}=0,\omega=0;T)$ (blue) in units of
eV$^{-1}$Fe$^{-1}$ vs. $T$ in Kelvin for a doping of $x=0$ (a) and
$x=0125$ (b). The interaction parameters have been chosen to be
$U=1.1$, $J'=0.25U$ (red) and $J=J'=0$, $U=V=1.35$ (a) or
$U=V=1.5$ (b), respectively (green).} \label{fig:ChiTx0}
\end{figure}

We now plot in Fig.~\ref{fig:ChiTx0} the
temperature dependence of both $\chi_0$ and the RPA enhanced
$\chi$ for the undoped and doped cases, for a particular choice of
interaction parameters. It is seen that in both the doped and
undoped cases, the susceptibility indeed increases
quasilinearly over the experimentally interesting range of 100 to
500K. This qualitative result appears to be relatively independent
of the choice of interaction parameters since it arises from the
band structure. The RPA enhancement for the cases shown is of
order 30\%.  While this factor increases as one approaches the
critical $U$ at which the Stoner instability is reached, it cannot
increase dramatically, because for the electronic structure of the
current model, the instability at large $q$ occurs first. Thus a large
enhancement (Ref.~\onlinecite{ref:Mazin}) of the $q=(0,0)$ susceptibility
does not occur in this model. We conclude
that the observed susceptibility contains a significant
temperature-independent interband or van Vleck component which is
not included in the $\chi_S$ calculated here.  Similar conclusions
were reached by the authors of Ref.~\onlinecite{ref:Grafe}.

\begin{figure}
\includegraphics[width=1\columnwidth]{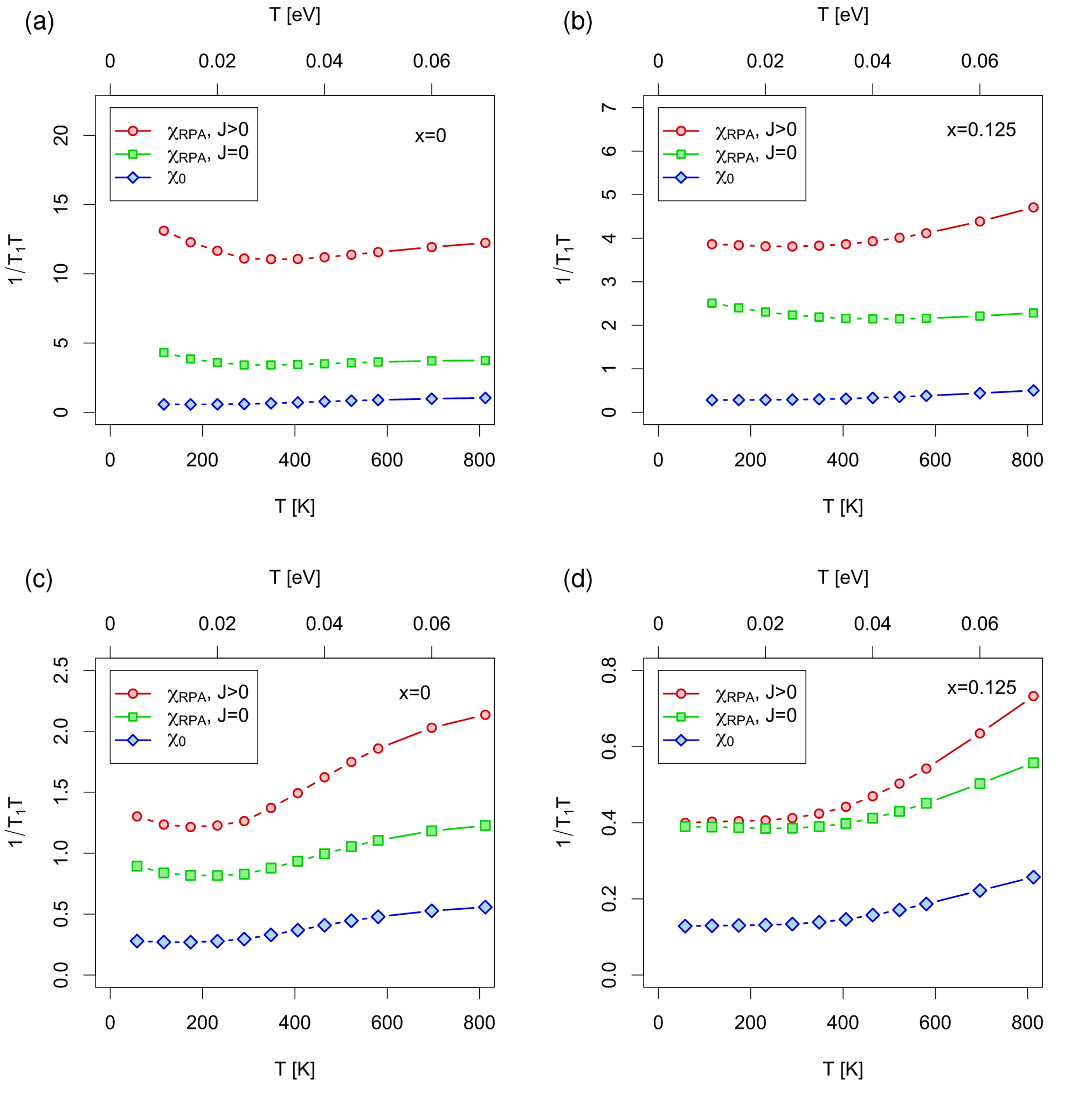}
\caption{(Color online) The NMR relaxation rate in units of
eV$^{-2}$Fe$^{-1}$ as a function of temperature $T$ for two
different dopings and with the same set of interaction parameters
and line colors as in Fig.~\ref{fig:ChiTx0}. Panels (a) and (b)
represent $x=0$ and $x=0.125$, respectively, for form factor
$A({\bf q})=1$. Panels (c) and (d) represent $x=0$ and $x=0.125$
for $A({\bf q})=\cos q_x/2 \cos q_y/2$.  } \label{fig:T1T}
\end{figure}

We also calculated the NMR relaxation rate $1/(T_1T)$ that is proportional to
to the imaginary part of the local ($q$ integrated) spin susceptibility
\begin{equation}
\frac{1}{T_1 T} = \frac{1}{N} \sum_q |A(q)|^2 \frac{\mathrm{Im} [\chi(q,\omega_L,T)]}{\omega_L}
\end{equation}
where $A(q)$ is a geometrical structure factor and $\omega_L$ is
the NMR frequency. In Fig~\ref{fig:T1T} we show $1/(T_1 T)$ as a
function of temperature for the same set of parameters that we
have used for the $q=0$ susceptibility. Here we find for the
interacting system an initial  decrease of the relaxation rate as
a function of temperature that is most significant for the undoped
compound with finite values of $J$. In the latter case we find a
very distinct upturn around 300 K followed by a slight and nearly
linear increase. For the other cases we find a similar tendency,
although not as pronounced.  In panels ~\ref{fig:T1T} (a) and (b)
we take the hyperfine form factor of the $^{19}$F nucleus used
e.g. in Ref. \onlinecite{ref:Ahilan} to be constant, since the F
is located directly above the Fe in the LOFFA material.  In panels
(c) and (d), we have modelled the form factor of the $^{75}$As
nucleus as $A({\bf q})=\cos q_x/2 \cos q_y/2 $, which suppresses
the ($\pi,0$) fluctuations, since the As atom is in the center of
a square containing four Fe.

Figures \ref{fig:ChiTx0} and \ref{fig:T1T} are intended to show
some of the dependence of the susceptibility and $T_1^{-1}$ on
interaction parameters for a few cases.  The trends are
qualitatively similar to
experiments\cite{ref:Ahilan,ref:Grafe,ref:RKlingeler}, although
the increase with increasing  temperature  for both quantities
found here is somewhat weaker than observed.  We have not
attempted to fit experiments in detail.

\section{The effective interaction and the gap function}
\label{sec:5}

\begin{figure}
\includegraphics[width=.8\columnwidth]{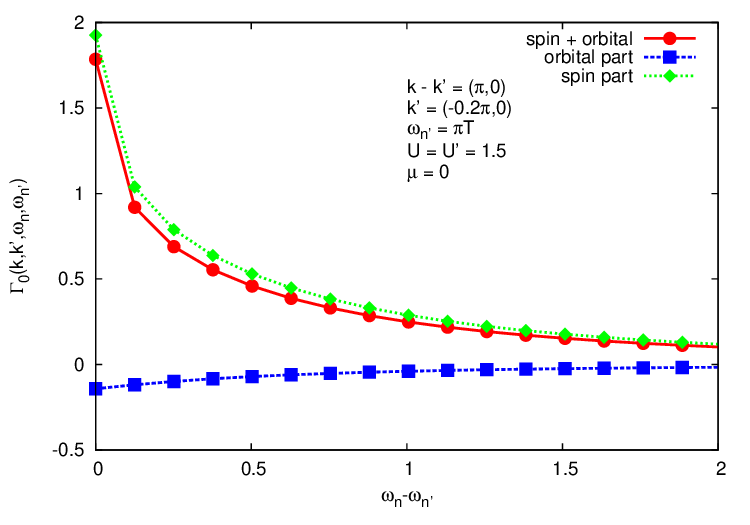}
\caption{(Color online) The energy dependence of the spin and orbital part of
the effective pairing interaction $\Gamma(k,k',\omega_n,\omega_n')$.}
\label{fig:Gammawndep}
\end{figure}

Assuming that the pairing interaction responsible for the
occurrence of superconductivity in the iron pnictides arises from
the exchange of spin and charge fluctuations, we can calculate the
pairing vertex using the fluctuation exchange
approximation~\cite{ref:Bickers}. For the multi-orbital
case~\cite{ref:Takimoto}, the singlet pairing vertex is given by
\begin{eqnarray}
&&{\Gamma}_{st}^{pq} (k,k',\omega) = \left[\frac{3}{2} U^s
\chi_1^{RPA}  (k-k',\omega) U^s + \nonumber \right.\,~~~~~~\,\\
&&\,~~~~~\left.
 \frac{1}{2} U^s
 - \frac{1}{2} U^c  \chi_0^{RPA}  (k-k',\omega)
U^c + \frac{1}{2} U^c \right]_{ps}^{tq}
\label{eq:fullGamma}
\end{eqnarray}
The $\chi_1^{RPA}$ term describes the spin fluctuation contribution
and the $\chi_0^{RPA}$ term the orbital (charge) fluctuation
contribution. Fig.~\ref{fig:Gammawndep} shows the Matsubara
frequency dependence of these two terms for a momentum transfer
$(\pi,0)$ and a typical interaction strength. Here one sees the that the
spin fluctuation contribution is dominant and falls off on a frequency
scale which is small compared with the bandwidth. Thus while the
gap equation depends upon the kernel $\mathrm{Im}
\Gamma_{st}^{pq} (k,k',\omega))$, the important $k$ and $k'$ values
are restricted by this frequency cut-off to remain near the Fermi surfaces.
Then, just as for the
electron-phonon case, the strength of the pairing interaction is
characterized by a frequency integral of this kernel weighted by
$\omega^{-1}$. Making use of the Kramers-Kronig relation
\begin{equation}
\int_0^\infty d\omega \frac{\mathrm{Im}\left[{\Gamma}_{st}^{pq} (k,k',\omega)\right]}{\pi \omega}
= \mathrm{Re}\left[{\Gamma}_{st}^{pq} (k,k',\omega=0)\right]
\end{equation}
we can proceed further by considering only the real part of the
$\omega=0$ pairing interaction.
If we now confine our considerations to the vicinity of the
Fermi surfaces we can determine the scattering of a Cooper
pair from the state $(k,-k)$ on the Fermi surface $C_i$ to the state
$(k',-k')$ on the Fermi surface $C_j$ from the projected interaction
vertex
\begin{eqnarray}
{\Gamma}_{ij} (k,k') & = & \sum_{stpq} a_{\nu_i}^{t,*}(-k)  a_{\nu_i}^{s,*}(k)
\mathrm{Re}\left[{\Gamma}_{st}^{pq} (k,k',0) \right] \nonumber \\
&& \times a_{\nu_j}^{p}(k')  a_{\nu_j}^{q}(-k')
\end{eqnarray}
where the momenta $k$ and $k'$ are restricted to the different
Fermi surface sheets with $k \in C_i$  and $k' \in C_j$. If we
decompose the superconducting gap into an amplitude $\Delta$ and a
normalized symmetry function $g(k)$ we can define a dimensionless
pairing strength functional\cite{ref:ScalapinoLohHirsch} as
\begin{equation}
\lambda [g(k)] = - \frac{\sum_{ij} \oint_{C_i} \frac{d k_\parallel}{v_F(k)} \oint_{C_j}
\frac{d k_\parallel'}{v_F(k')} g(k) {\Gamma}_{ij} (k,k')
g(k')}{ (2\pi)^2 \sum_i \oint_{C_i} \frac{d k_\parallel}{v_F(k)} [g(k)]^2 }
\end{equation}
 Here $\Gamma_{ij}$ is only the symmetric part
\begin{equation}
\frac{1}{2} \left[ {\Gamma}_{ij} (k,k') + {\Gamma}_{ij} (k,-k') \right]
\end{equation}
of the full interaction, which gives identical results within the
spin singlet subspace.  The Fermi velocity is defined to be
$v_{F}(k) = |\nabla_k E_\nu(k)|$ for $k$ on the given Fermi
surface. From the stationary condition we find the following
eigenvalue problem
\begin{equation}
- \sum_j \oint_{C_j} \frac{d k_\parallel'}{2\pi} \frac{1}{2\pi v_F(k')} {\Gamma}_{ij} (k,k')
g_\alpha(k') = \lambda_\alpha g_\alpha(k) \label{EVP}
\end{equation}
The kernel $\Gamma_{ij}(k,k')$ is evaluated at temperatures
well below the characteristic temperature at which the spin
fluctuation spectrum has formed. In this temperature region, the
interaction is independent of temperature. From the above
eigenvalue problem we will determine the leading eigenfunction
$g_\alpha (k)$ and eigenvalue $\lambda_\alpha$ for a given
interaction vertex $\Gamma_{ij}(k,k')$. The largest eigenvalue
will lead to the highest transition temperature and its
eigenfunction determines the symmetry of the gap. The next leading
eigenvalues and eigenfunctions further characterize the pairing
interaction and can indicate the structure of possible collective
states. With the knowledge of the pairing function $g_\alpha(k)$
we can also determine the individual contributions of the
different intra- and interorbital scattering processes to the
total pairing strength $\lambda_\alpha$.

\section{Results for the pairing strength $\lambda$ and the gap function $g(k)$}
\label{sec:6}

\subsection{Eigenvalue problem}

In this section, we  present and discuss results for the pairing
strength $\lambda$ and the symmetry function $g(k)$ obtained
within our weak-coupling approach.

Solving the eigenvalue problem given in Eq.~\ref{EVP}, we find a
set of eigenfunctions $g_\alpha(k)$ defining the $k$ dependence of
the gap on the FS sheets together with a set of corresponding
eigenvalues $\lambda_\alpha$ denoting the dimensionless pairing
strengths associated with a given $g_\alpha(k)$. We first have to
classify the different eigenfunctions according to basic symmetry
operations. In the following we will speak of an $s$ wave state if
\begin{eqnarray}
g(-k_x,k_y) & = & g(k_x,-k_y) = g(k_x,k_y) \nonumber \\
g(k_y, k_x) & = & g(k_x,k_y)
\end{eqnarray}
while we have a $d_{x^2-y^2}$ wave state if
\begin{eqnarray}
g(-k_x,k_y) & = & g(k_x,-k_y) = g(k_x,k_y) \nonumber \\
g(k_y, k_x) & = & -g(k_x,k_y)
\end{eqnarray}
Since there are two different $d_{x^2-y^2}$ wave states among the
first few eigenfunctions, we have to distinguish them further,
e.g. by comparing the sign of $g_\nu(k)$ for $\nu=\alpha_1$ and
$\nu=\alpha_2$ in the same direction in momentum space. Here we
will label the state that changes sign between $\alpha_1$ and
$\alpha_2$ with $d_{x^2-y^2}(1)$, the one without sign change with
$d_{x^2-y^2}(2)$. Furthermore we can distinguish a $d_{xy}$ wave
state that is given by
\begin{eqnarray}
g(-k_x,k_y) & = & g(k_x,-k_y) = - g(k_x,k_y) \nonumber \\
g(k_y, k_x) & = & g(k_x,k_y)
\end{eqnarray}
and a $g$ wave state with
\begin{eqnarray}
g(-k_x,k_y) & = & g(k_x,-k_y) =- g(k_x,k_y) \nonumber \\
g(k_y, k_x) & = & -g(k_x,k_y)
\end{eqnarray}

\subsection{Results: $J=J'=0$}

\begin{figure}
\includegraphics[width=1\columnwidth]{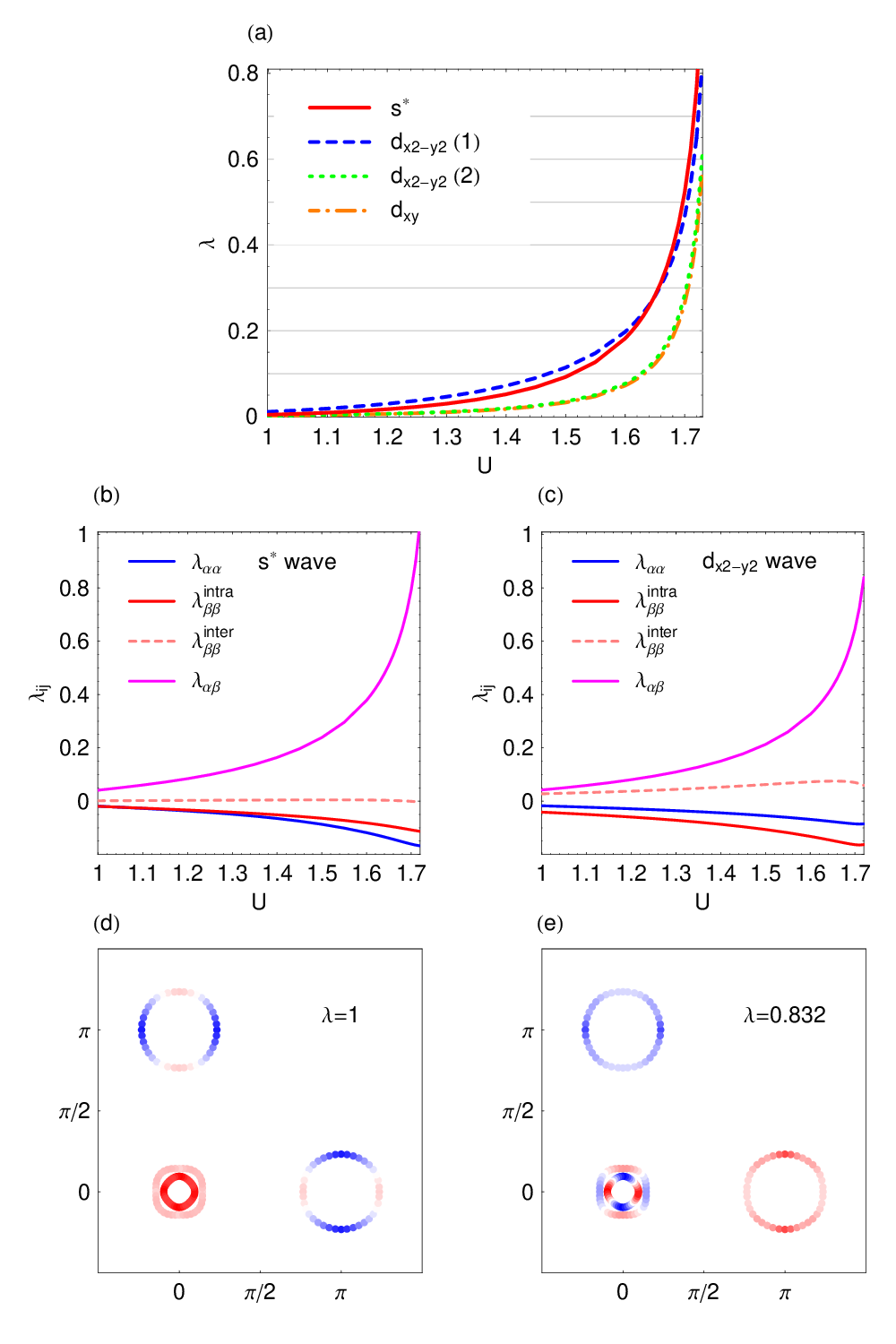}
\caption{(Color online) The eigenvalues and eigenfunctions for the
electron-doped compound ($x=0.125$)  for $U=V$ and $J=J'=0$.
The four largest eigenvalues as a function of $U$ (a) and the different
inter- and intraband contributions to the eigenvalues $\lambda$ for the
two symmetries with largest eigenvalues, extended $s$ (b) and $d_{x^2-y^2}$ (c)
wave. Color coded plot of the extended $s$ wave (d) and the $d_{x^2-y^2}$
wave (e) pairing functions along the different Fermi surface sheets,
calculated close to the instability ($U=V=1.73$).}
\label{fig:LambdaUx0125_Jp0}
\end{figure}

\begin{figure}
\includegraphics[width=1\columnwidth]{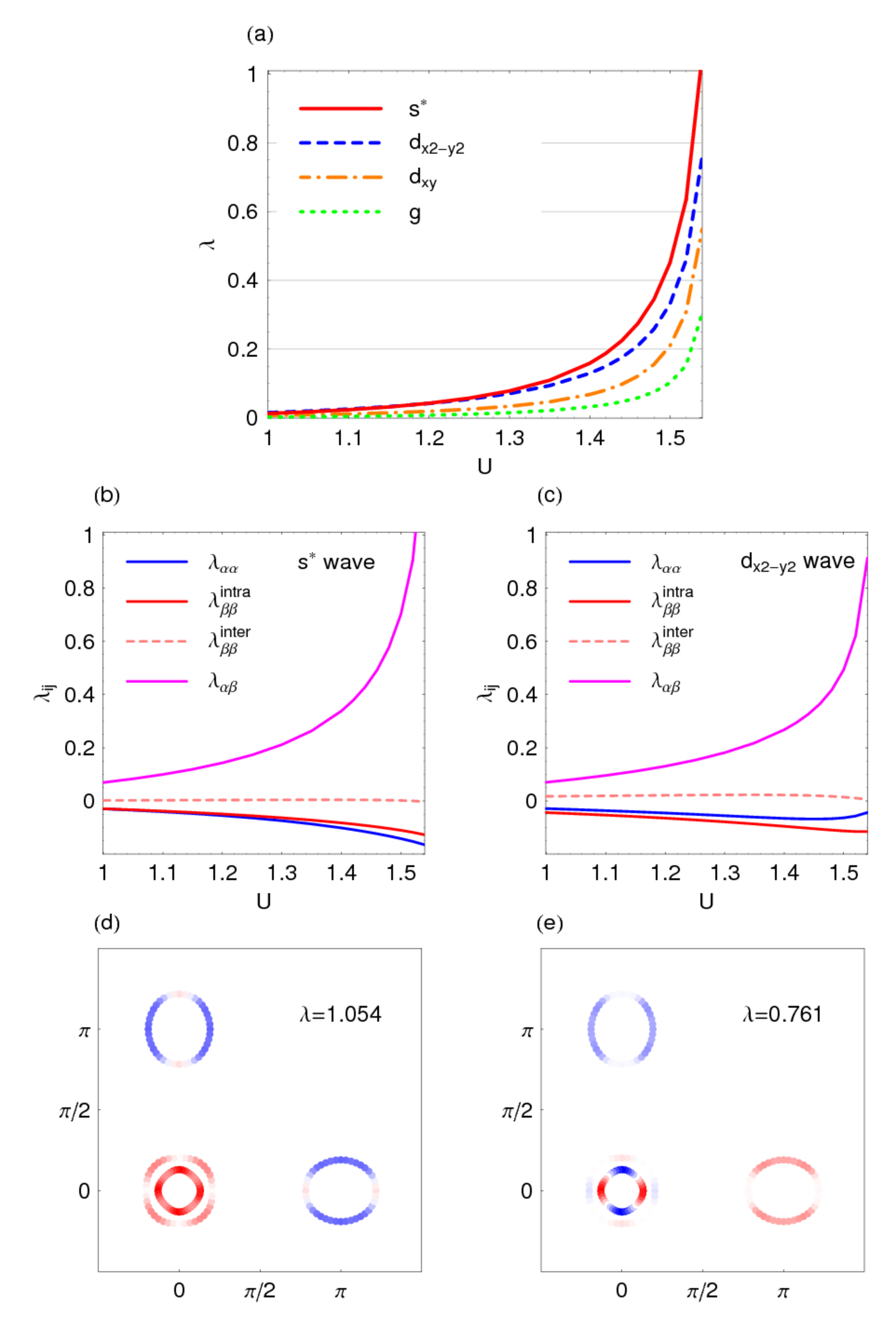}
\caption{(Color online) The same as in Fig.~\ref{fig:LambdaUx0125_Jp0}
but for the undoped compound ($x=0$) with $V=U$ and $J=J'=0$.
Here the two pairing functions shown in (d,e) correspond to the two leading
eigenvalues and are calculated for $U=1.54$.}
\label{fig:LambdaUx0_Jp0}
\end{figure}

First we consider a case where the Hund's rule coupling and the
pair hopping energy are negligible compared to the intra- and
interorbital Coulomb interactions, so we set $J=J'=0$ and $V=U$.
In Fig.~\ref{fig:LambdaUx0125_Jp0} (a) we show the pairing
strength eigenvalues $\lambda_\alpha$ for the four leading
eigenvalues as a function of $U$ for the electron-doped compound
($x=0.125$). Approaching the critical value of $U$ where the
eigenvalues start to diverge we find a clear separation of the two
leading eigenvalues from the next two eigenvalues. However the two
leading eigenvalues, corresponding to the $s$ and the
$d_{x^2-y^2}$ symmetry remain very similar in size and we find a
crossover from the $d_{x^2-y^2}$ to the $s$ symmetry around
$U=1.65$.
We also show the $s$ wave (d) and the $d_{x^2-y^2}$ wave (e)
pairing functions on the four FS sheets for $U=1.73$ close to its
critical value\cite{footnote1}. The extended $s$ wave state (which we have
labelled $s$) is characterized by i) a sign change on the $\beta$
FS sheets with nodal points displaced from the generic
$(0,\pi)$-$(\pi,0)$ direction that would result from a pure $\cos
k_x + \cos k_y$ state, and ii) a change of average sign on the
$\beta$ sheet relative to that on the $\alpha$ sheets. On the
$\alpha$ FS sheets we find a nodeless but anisotropic gap
distribution with higher weight on the small $\alpha_1$ compared
to the larger $\alpha_2$ sheet. The $d_{x^2-y^2}$ state features
an anisotropic gap distribution on the $\beta$ FS sheets and a
rather conventional $d$ wave gap distribution on the $\alpha$ FS
surfaces, with a sign change between the $\alpha_1$ and $\alpha_2$
sheet. In Fig.~\ref{fig:LambdaUx0_Jp0} we show the corresponding
results for the undoped ($x=0$) compound. Here we find similar
results, although the eigenvalues close to the instability are
better separated and the crossover from $d_{x^2-y^2}$ to $s$
appears already for a rather small value of $U$.
If we compare the $s$ wave symmetry function for $x=0$ close to
the instability (see Fig.~\ref{fig:LambdaUx0_Jp0} (d)) to the
corresponding symmetry function for the electron-doped compound,
we find that the nodal points on the $\beta$ FS have moved even
closer to the tips of the sheets and the negative dip between them
has decreased considerably. For the $d_{x^2-y^2}$ symmetry (see
Fig.~\ref{fig:LambdaUx0_Jp0} (e) we find that the weight on the
$\alpha_2$ FS sheet has nearly vanished.

In Figs.~\ref{fig:LambdaUx0125_Jp0} and \ref{fig:LambdaUx0_Jp0},
as well as in the following Figs.~\ref{fig:LambdaUx0125_Jp025}
and \ref{fig:LambdaUx0_Jp025} we compare in panels (b) and (c) 
the contributions of the different
intra- and interband processes. Here $\lambda_{\alpha \alpha}$ sums up all
contributions of $\lambda_{ij}$ resulting from scattering within each
$\alpha$ sheets and in between the two $\alpha$ sheets as
$\lambda_{\alpha\alpha} = \sum_{i,j=1,2} \lambda_{ij}$
where $i=1$ refers to the $\alpha_1$, $i=2$, to the  $\alpha_2$, $i=3$
to the $\beta_1$ and $i=4$ to the $\beta_2$ Fermi surface.
$\lambda_{\beta \beta}^{intra} =  \lambda_{33} + \lambda_{44}$
denotes the intraband contributions of the $\beta$ FS sheets, while
$\lambda_{\beta \beta}^{inter} = \lambda_{43} + \lambda_{34}$
 is the corresponding interband contribution. Finally
$\lambda_{\alpha \beta}$ sums up all the remaining contributions,
resulting from scattering between the $\alpha$ and $\beta$ Fermi
surfaces. It is obvious that in the case of $J=J'=0$ for both of
the main pairing symmetries the intersheet $\lambda_{\alpha
\beta}$ contribution is responsible for the rapid increase of the
pairing strength $\lambda$ with $U$, reflecting the  nesting
between the $\alpha$ and $\beta$ FS sheets with nesting vector
$Q^*$. All other contributions are small and mainly negative. For
finite values of $J$ and $J'$, which we will next discuss we find
that for the $d_{x^2-y^2}$ pairing symmetry the interorbital
contributions between the two $\beta$ sheets become also
important and for the electron-doped compound even dominant.

\subsection{Results: $J,J'>0$}

\begin{figure}
\includegraphics[width=1\columnwidth]{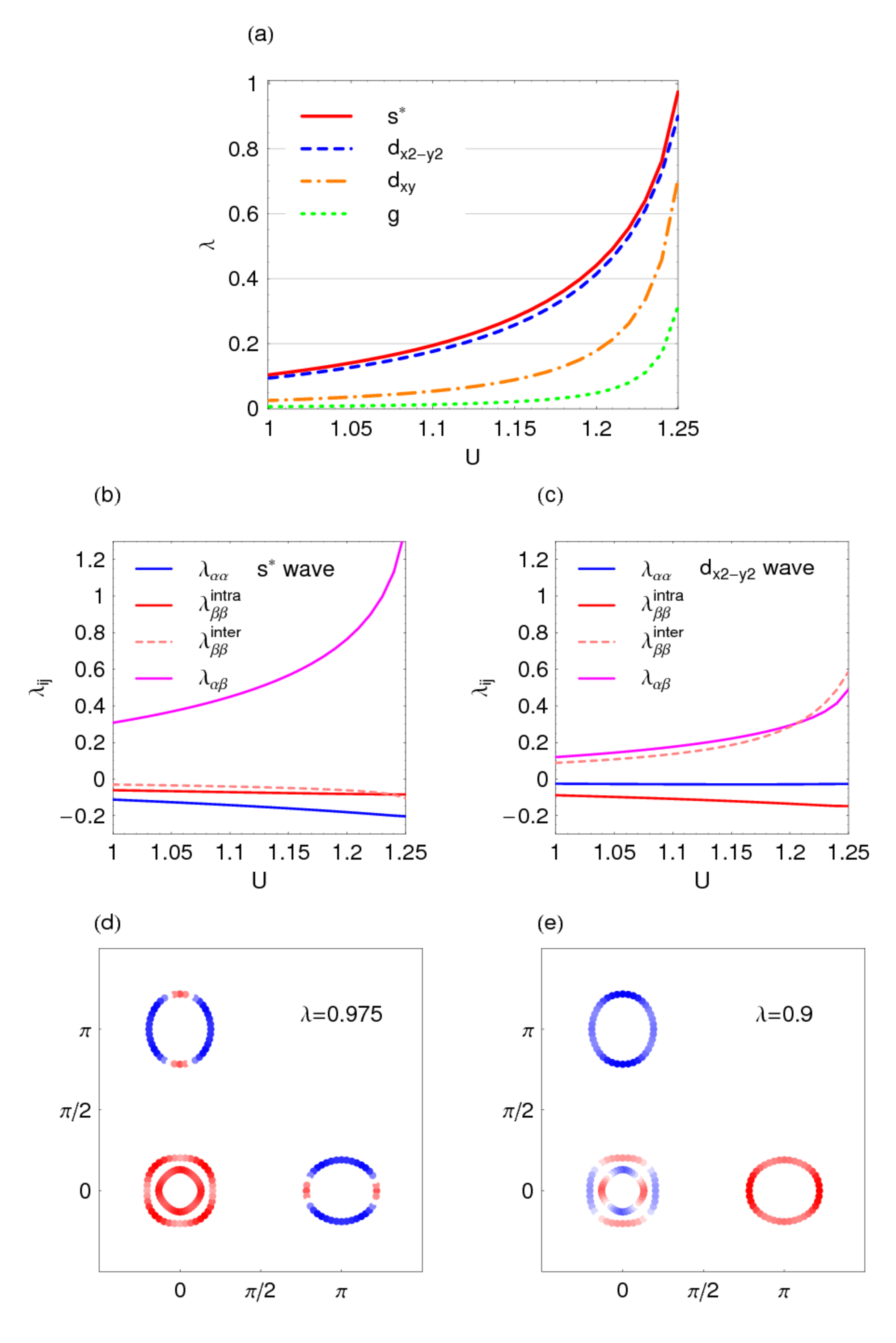}
\caption{(Color online) The same as in Fig.~\ref{fig:LambdaUx0125_Jp0}
for the undoped compound ($x=0$) with $J'=J/2=U/4$ and $V=U-5/4J$.
Again the extended $s$ wave (d) and the $d_{x^2-y^2}$ wave (e)
pairing functions are calculated close to the instability with $U=1.25$.}
\label{fig:LambdaUx0_Jp025}
\end{figure}

\begin{figure}
\includegraphics[width=1\columnwidth]{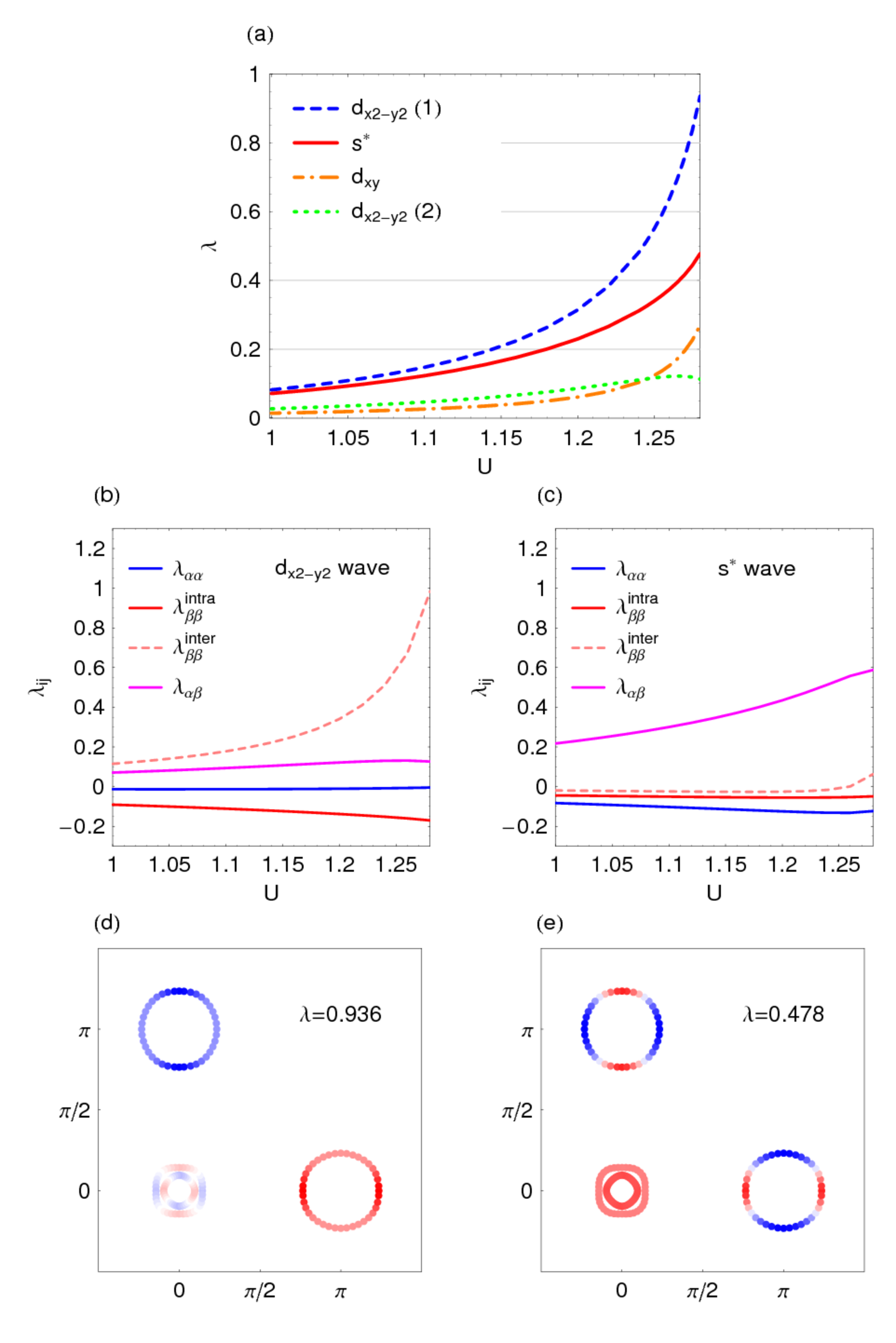}
\caption{(Color online) The same as in Fig.~\ref{fig:LambdaUx0125_Jp0}
for the electron-doped compound ($x=0.125$) with $J'=J/2=U/4$ and $V=U-5/4J$.
Here the $d_{x^2-y^2}$ wave (d) is more stable than the
extended $s$ wave (e) pairing function and the corresponding eigenvalues are
clearly separated. Both are calculated close to the instability with $U=1.28$.}
\label{fig:LambdaUx0125_Jp025}
\end{figure}

Now we  consider a case of finite $J$ and $J'$. Here we will
choose $J'=J/2$ and we will fix $V=U-3/4J-J'$.  These choices are
consistent with the generation of intrasite couplings from a
Hubbard argument~\cite{ref:oles,ref:kubo}, as mentioned above, as
well as with the range of values for $J/U$ found by Anisimov {\it
et al.} in an {\it ab initio} calculation~\cite{ref:anisimov}.  In
Fig.~\ref{fig:LambdaUx0_Jp025} (a) we show for the undoped
compound ($x=0$) and for $J'=J/2=U/4$ the same four eigenvalues as
a function of $U$ as in the previous figures. Here we find again
that the extended $s$ wave state (d) is the most stable pairing
configuration followed by the $d_{x^2-y^2}$ wave state (e). For
the $d_{x^2-y^2}$ state the main contributions to the pairing
function are along the $\beta$ FS sheets, while the $\alpha$ FS
sheets contribute less significantly. Since this state has no
nodes along the $\beta$ sheets, it can be considered as a mainly
nodeless state, in contrast to the states found for $J=J'=0$. The
same is true for the electron-doped case with $x=0.125$
(Fig.~\ref{fig:LambdaUx0125_Jp025}). Here the  $d_{x^2-y^2}$ state
is the most stable state followed by the extended $s$ wave state.
Close to the instability the nodeless $\beta$ sheets contribute
dominantly to the pairing.

\begin{figure}
\includegraphics[width=1\columnwidth]{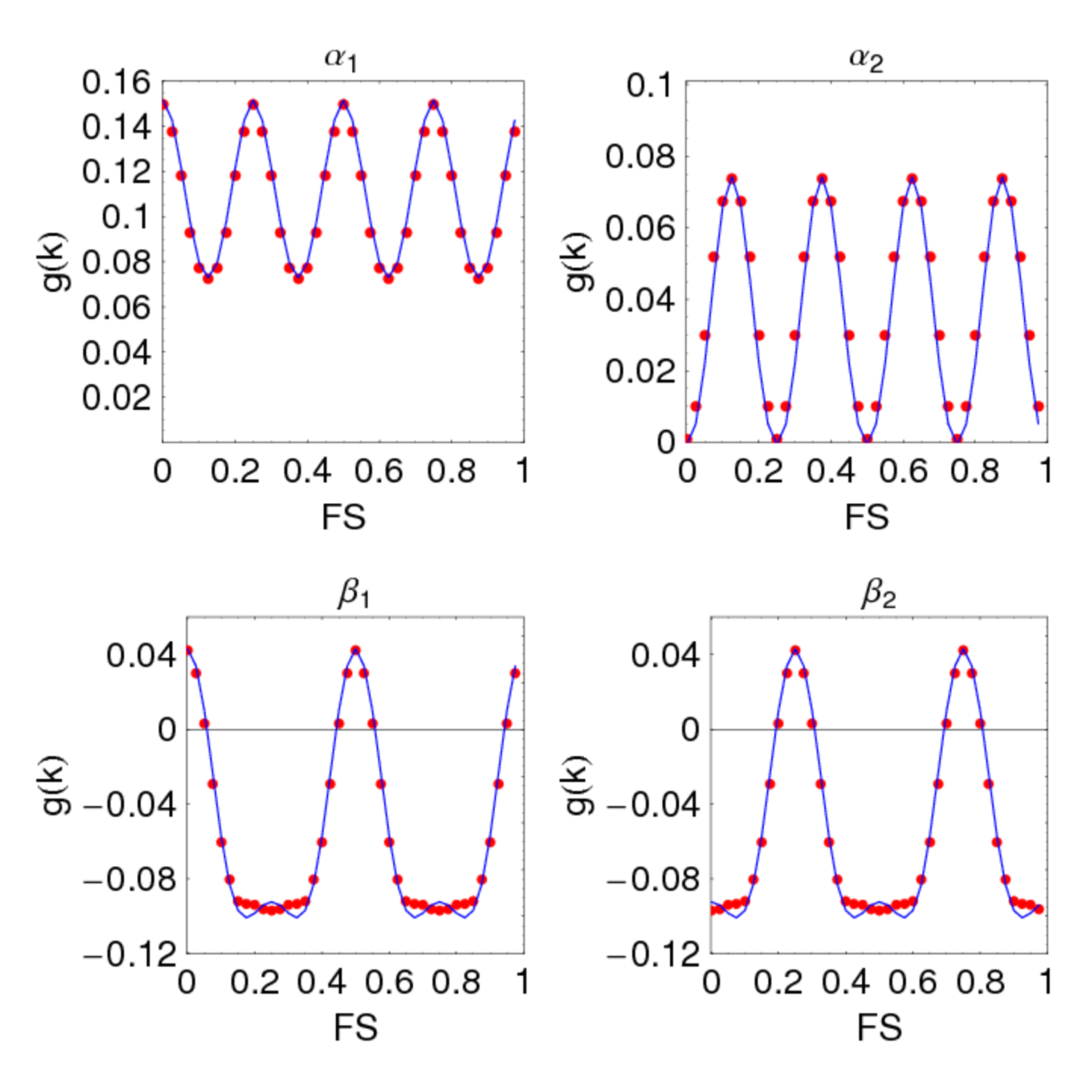}
\caption{(Color online) The $s$ wave symmetry function $g(k)$ on the different
FS sheets for $U=1.5$ and $x=0$. Here we compare a fit of $g_\nu(k)$ (blue line)
to the actually calculated values of $g(k)$ (red points).}
\label{fig:FitgkSwave}
\end{figure}

\begin{figure}
\includegraphics[width=1\columnwidth]{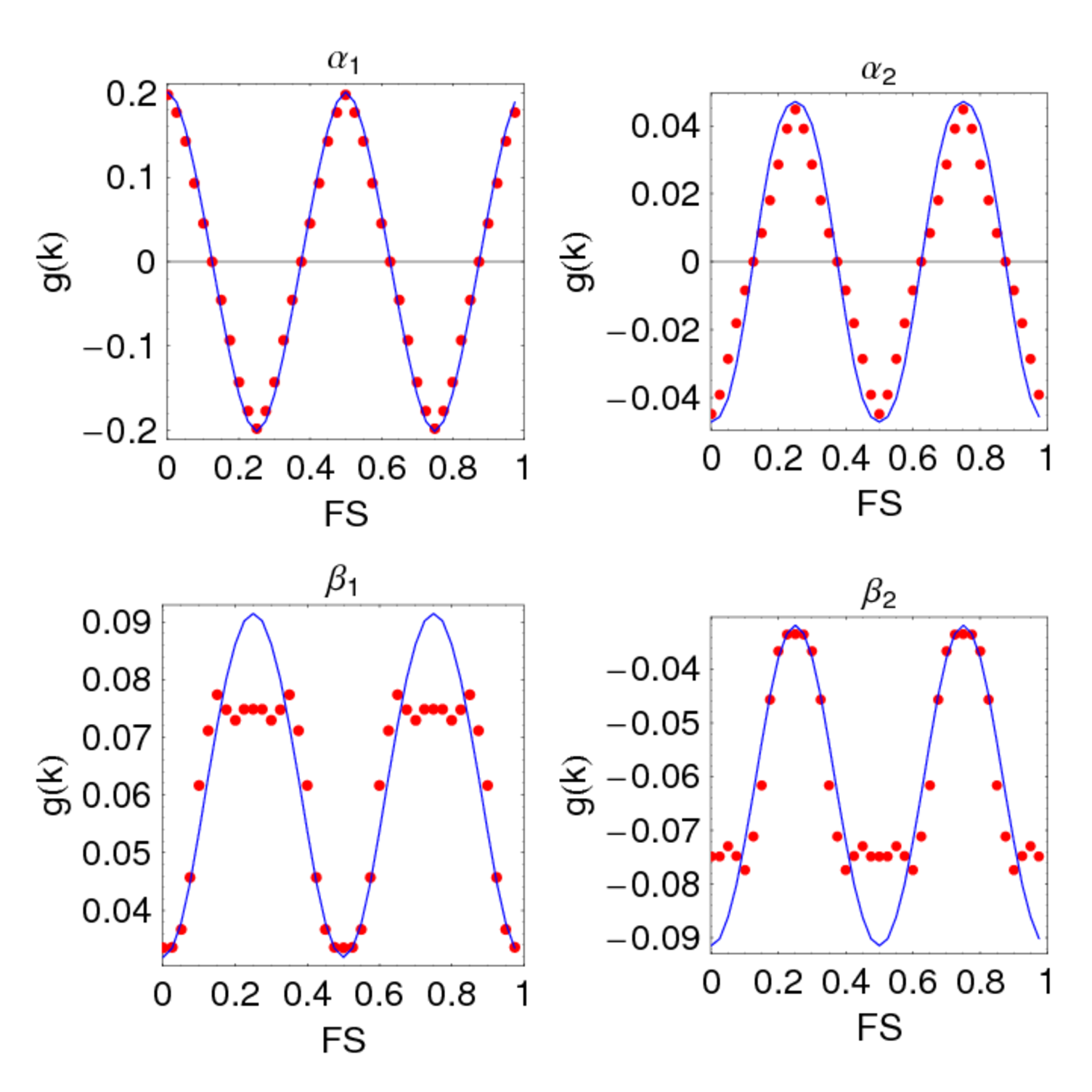}
\caption{(Color online) The $d$ wave symmetry function $g(k)$ on the different
FS sheets for $U=1.5$ and $x=0$. Here we compare a fit of $g_\nu(k)$ (blue line)
to the actually calculated values of $g(k)$ (red points).}
\label{fig:FitgkDwave}
\end{figure}

Although the pairing is constrained to the close vicinity of the
Fermi surfaces, one can try to find an approximation of the
$g_\nu(k)$ on the different FS sheets that extends further into
the Brillouin zone. Here we want to restrict ourselves to the
smallest number of harmonics necessary to find a reasonable fit of
the pairing function on the different FS sheets. For the $s$ wave
symmetry we can write
\begin{eqnarray}
g_\nu(k) & = & 2 A_\nu \left[ \cos k_x +\cos k_y + 2 w_{\nu,xy} \cos k_x \cos k_y \right. \nonumber \\
& & \left. + 2 w_{\nu,4x4y} \cos 4 k_x \cos 4 k_y \right]
\end{eqnarray}
where we find for $U=1.5$ the following parameters: $A_{\alpha_1}
= 0.051$, $w_{\alpha_1,xy} = -0.35$, and $w_{\alpha_1,4x4y} =
-1.35$ for the $\alpha_1$ sheet, $A_{\alpha_2} = 0.02$,
$w_{\alpha_2,xy} = -0.1$, and $w_{\alpha_2,4x4y} = 1$ for the
$\alpha_2$ sheet, and $A_{\beta} = 0.15$, $w_{\beta,xy} = 0.17$,
and $w_{\beta,4x4y} = -0.1$ for the $\beta$ sheets. In the same
way, we can find an approximation to the $d$ wave symmetry
function as
\begin{equation}
g_\nu(k)  =  2 A_\nu \left[ \cos k_x - \cos k_y +  w_{\nu,2x} (\cos 2 k_x - \cos 2 k_y) \right]
\end{equation}
with $A_{\alpha_1} = -1.2$ and $w_{\alpha_1,2x} = 0$
for the $\alpha_1$ sheet, $A_{\alpha_2} = 0.12$ and $w_{\alpha_2,2x} = 0$,
for the $\alpha_2$ sheet, and
$A_{\beta} = -0.018$ and $w_{\beta,2x} = -1.12$ for the $\beta$ sheets.
The results of the fitting are shown in Figs.~\ref{fig:FitgkSwave} and \ref{fig:FitgkDwave}.

\subsection{Role of nesting}

\begin{figure}
\includegraphics[width=0.7\columnwidth]{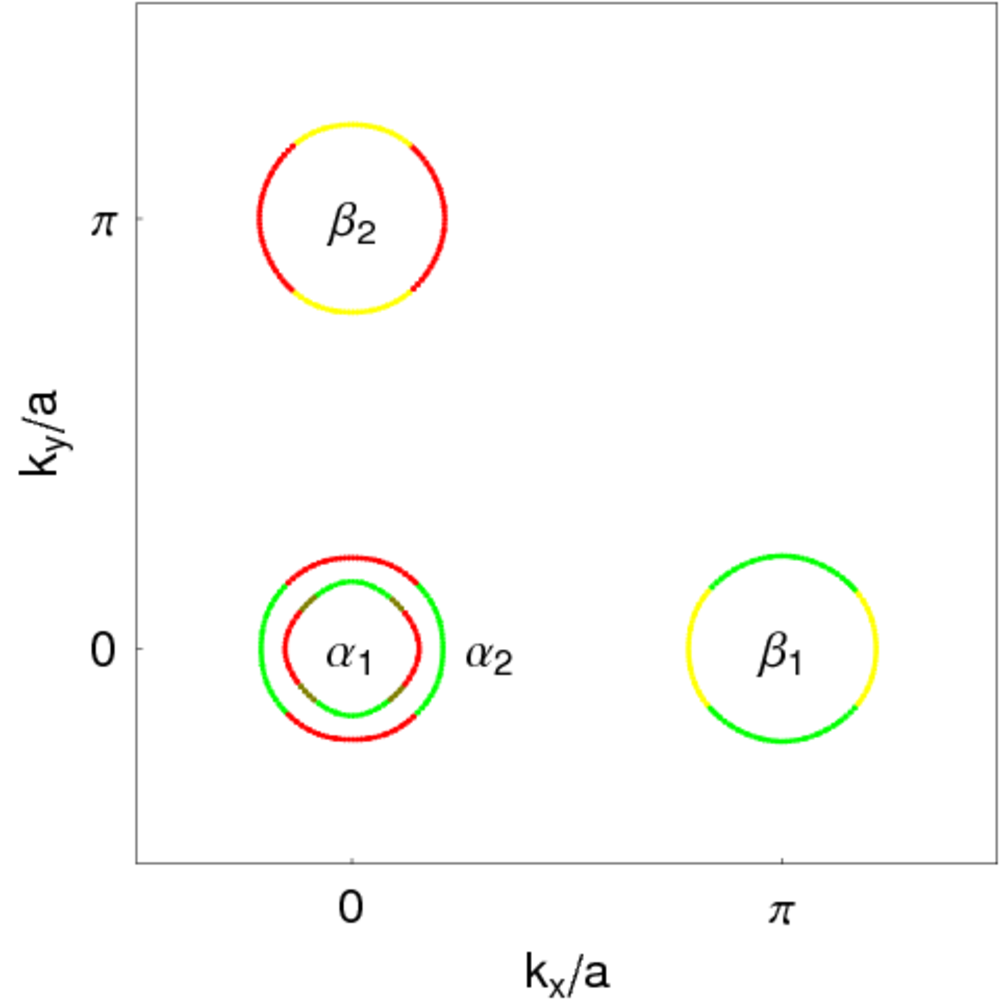}
\caption{(Color online) The FS sheets found within a circular approximation
showing nearly perfect nesting between the $\alpha_2$ and the $\beta$
FS sheets with nesting vectors $(\pi,0)$, or $(0,\pi)$, respectively.
Here we use the same color convention as in Fig.~\ref{fig:BSFS5band}.}
\label{fig:sphericalFS}
\end{figure}

To study the influence of nesting on the spin susceptibility,
especially on the $(\pi,0)$ peak, we slightly modify the hopping
parameters, creating a toy model with perfectly nested FS sheets.
This means that we try to find an approximation where the
$\alpha_2$ and the $\beta$ FS sheets are of approximately the same
size and shape. The band structure used within this toy model is
very similar to the band structure found formerly by a fit to the
DFT bands, to assure that all basic properties of the model are
reasonable, i.e. that the matrix elements are similar to the
correct matrix elements found from the DFT fit. In
Fig.~\ref{fig:sphericalFS} we show the near circular FS sheets of
this model.

\begin{figure}
\includegraphics[width=1.\columnwidth]{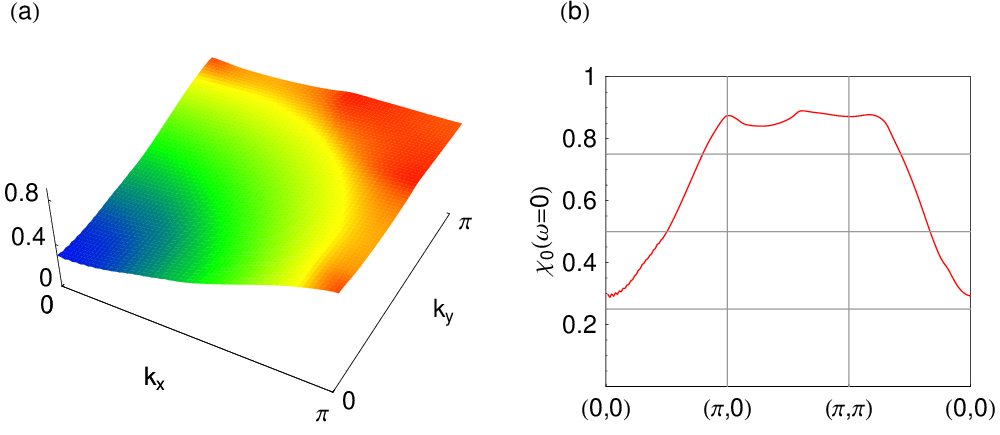}
\caption{(Color online) The susceptibility $\chi_0(q)$ for a toy
model with spherical FS sheets.} \label{fig:chi0spherical}
\end{figure}

\begin{figure}
\includegraphics[width=1.\columnwidth]{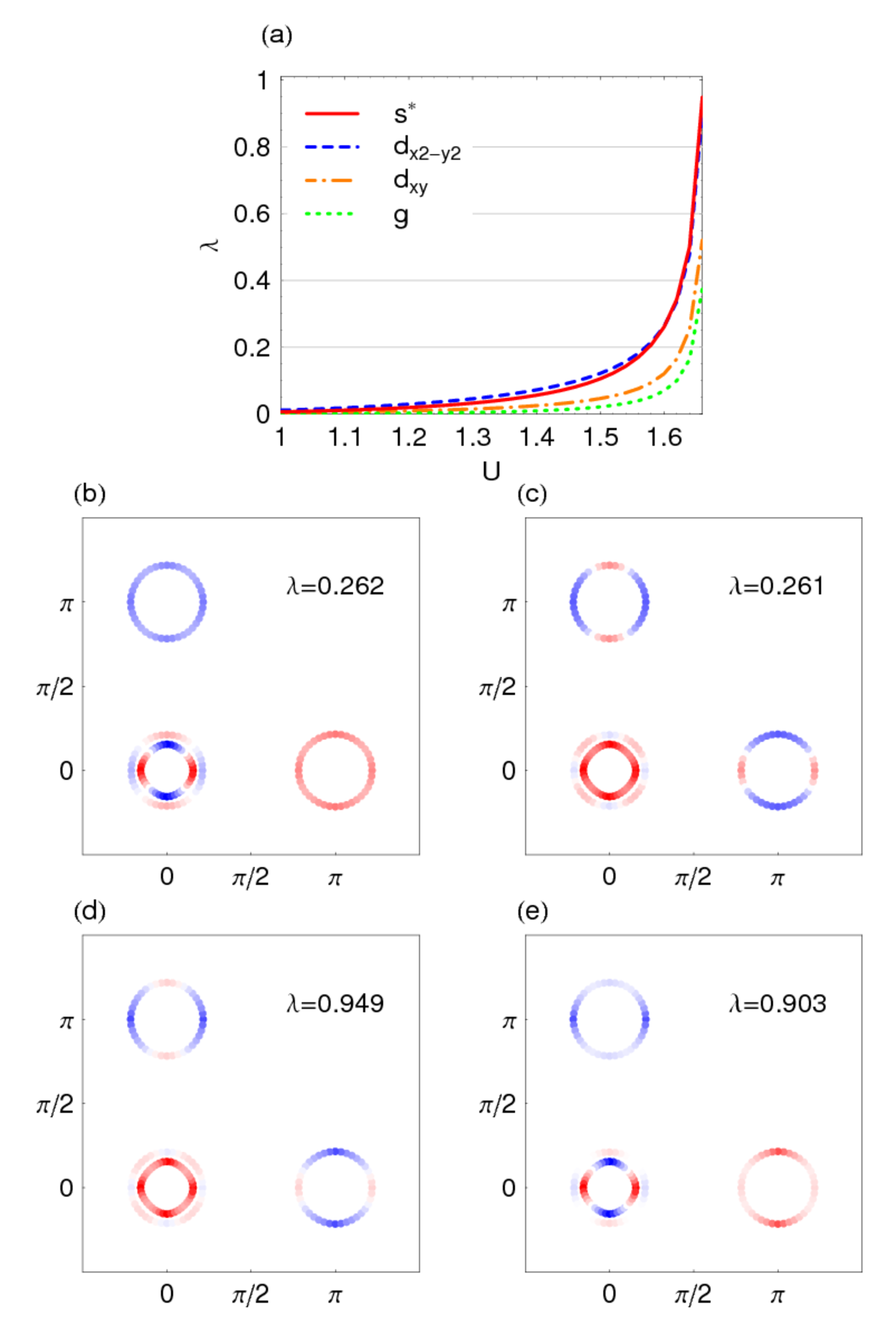}
\caption{(Color online) The leading eigenvalues $\lambda$ for the toy model
with spherical FS sheets (a). The two leading pairing functions $g(k)$ for the toy model
with spherical FS sheets for $U=V=1.6$ (b,c) and for $U=V=1.66$ (d,e).}
\label{fig:gkspherical}
\end{figure}

The bare spin susceptibility $\chi_0(q)$ for this model in the
undoped case is exhibited in Fig.~\ref{fig:chi0spherical}, and may
be compared to the results of the original model in
Fig.~\ref{fig:Chi0}. We see that the $(\pi,0)$ peak remains but is
not particularly enhanced by the enhanced nesting, implying that
the original model was already quite close to a nesting condition.
On the other hand, some subtle differences which are quite
interesting appear when we examine the pairing functions.  If we
consider an  even more  {\it Gedanken}-type model where the two
$\alpha$ sheets have degenerate radius $a$, which is also the same
as the two $\beta$ sheets, we see that the Fermi surface in the
1st effective Brillouin zone is invariant under a translation by
$(\pi,0)$. Under this transformation, simple extended-$s$ and
$d_{x^2-y^2}$ functions $\cos k_x +\cos k_y$ and $\cos k_x - \cos
k_y$ map into one another identically.  We therefore expect that
the pairing eigenvalues for $s$ and $d_{x^2-y^2}$ will become
 degenerate.  The toy model band structure considered here and
 shown in Fig.~\ref{fig:sphericalFS} is nearly the same as the
 {\it Gedanken} model, but has two slightly nondegenerate $\alpha$
 sheets. We see nonetheless in Fig.~\ref{fig:gkspherical}
 (b,c) and (d,e) that the two competing states are
 indeed nearly exactly degenerate.
 Fig.~\ref{fig:gkspherical} (a) exhibits the dependence
 of the leading eigenvalues on $U$.

%
%\subsection{Comparison with 2-band model}
%
%{\bf PH: DJS would like us to add some brief results for the
%pairing in the 2-band model for comparison.  We have a plot e.g.
%for pairing lambda's vs. U?  What else is necessary/interesting?}

\section{The spatial and orbital structure}
\label{sec:7}

The gap function $g_\nu(k)$ contains information on the spatial and orbital structure
of the pairs. In the previous section, we have calculated $g_\nu(k)$ and found that
its behavior on the Fermi surfaces can be fit by low order harmonics for both the
$s$ and $d$-wave eigenstates. Here we use these results to determine simple pictures
of the internal pair structure. A gap operator can be written as
\begin{equation}
\Delta = \frac{1}{N} \sum_{\nu,k} g_\nu(k) \gamma_{\nu\uparrow}(k)
\gamma_{\nu\downarrow}(-k)
\end{equation}
with $\gamma_{\nu\sigma}(k)$ the destruction operator for an electron in the
$\nu^{\mathrm th}$ band with wave vector $k$ and spin $\sigma$. Using the
band-orbital matrix elements to relate the band operator to the orbital operator
\begin{equation}
\gamma_{\nu\sigma}(k) = \sum_n a_\nu^n( k) d_{n\sigma}(k)
\end{equation}
we have
\begin{equation}
\Delta = \frac{1}{N} \sum_{l_1,l_2} a_{nm}(l_1 - l_2) d_{n\uparrow}(l_1)
d_{m\downarrow}(l_2)
\end{equation}
with
\begin{equation}
a_{nm}(l) = \frac{1}{N} \sum_{k \nu} g_\nu(k) a_\nu^n( k) a_\nu^m( -k)
e^{-i k l}
\end{equation}
where $l=l_1-l_2$. The amplitude $a_{nm}(l)$ describes the internal spatial
and orbital structure of the pair.

Using the harmonic approximations for $g_\nu(k)$ for the $s$ and
$d_{x^2-y^2}$ gaps discussed in the previous section, we have
calculated  $a_{nm}(l)$. In principle the $k$ sum should be
cut-off when $k-k_F$ exceeds the inverse coherence length. Here we
have used a Gaussian cutoff of the ${\bf k}$ sum which decays when
$|{\bf k} -{\bf k}_F|$ exceeds $2\pi/\xi_0$, where $\xi_0$ is
taken to be of order 3 times the Fe-Fe spacing $a$. This provides
a local picture of the internal orbital structure of a pair.  This
basic structure continues out to a radius set by the coherence
length $\xi_0$.

\begin{figure}
\includegraphics[width=1\columnwidth]{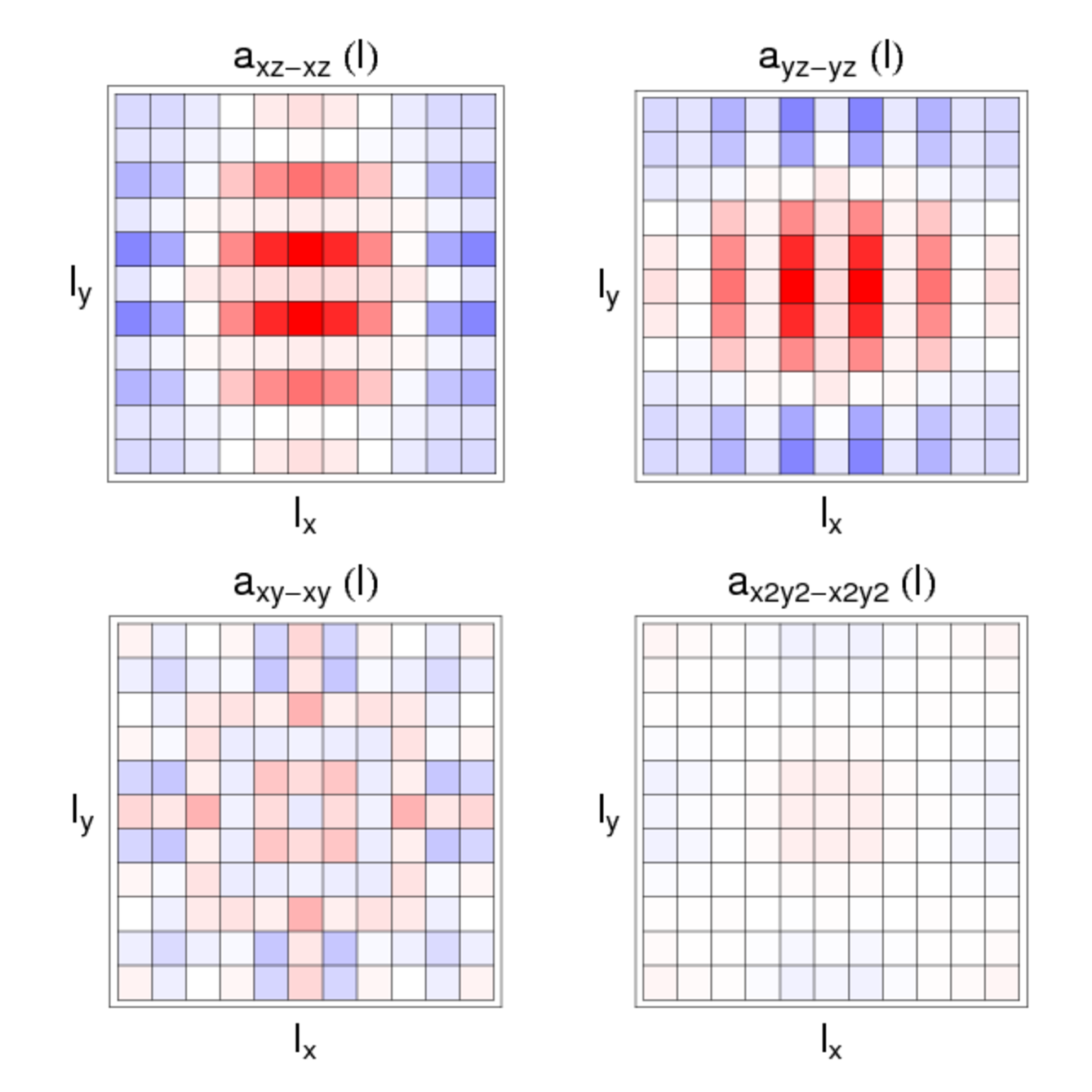}
\caption{(Color online) The spatial and orbital pair structure for
the $s$ gap calculated from the harmonic approximation of $g(k)$
corresponding to $U=1.5$ with a cut-off length of $\lambda_c = 3.3
d_{Fe-Fe}$.} \label{fig:spatials}
\end{figure}

\begin{figure}
\includegraphics[width=1\columnwidth]{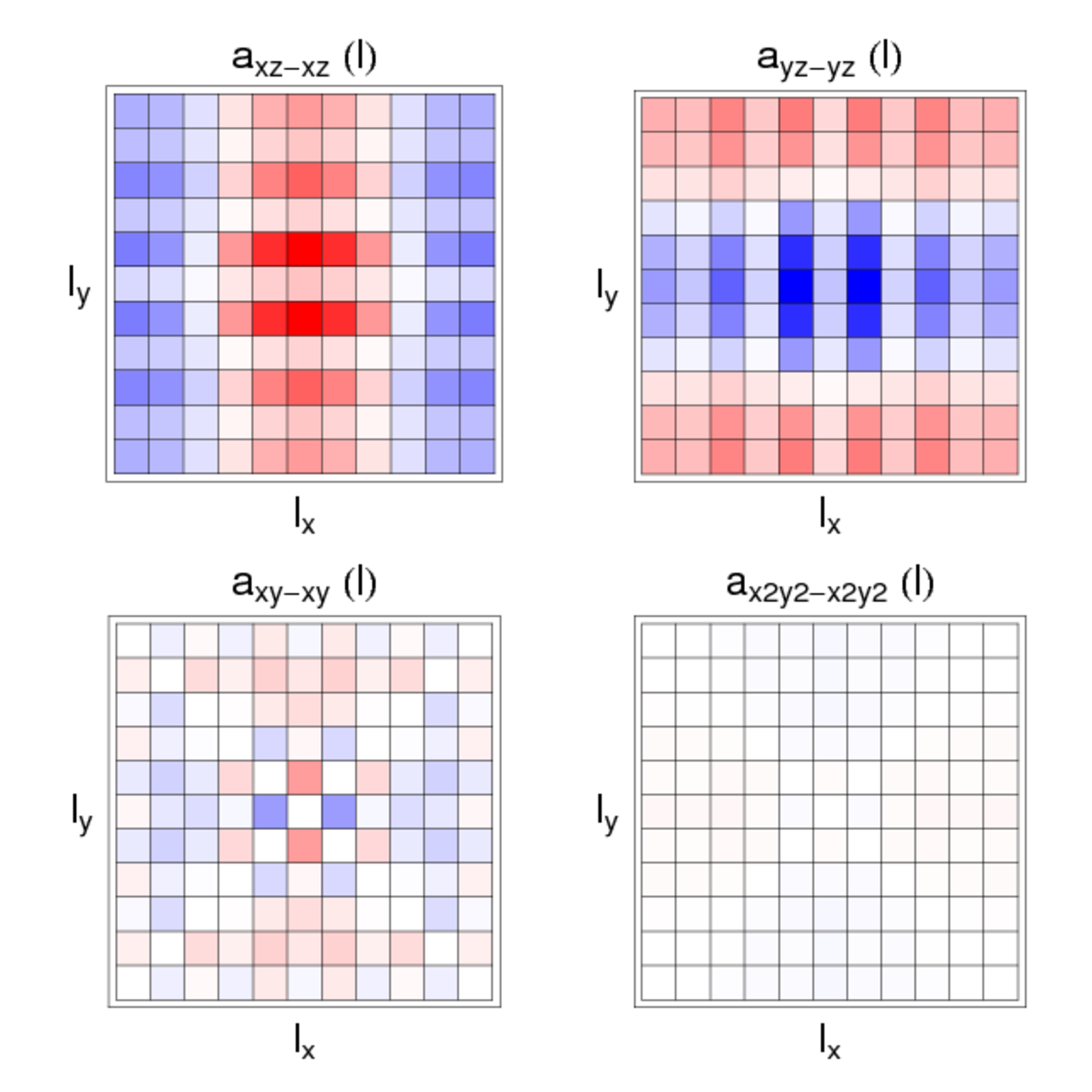}
\caption{(Color online) The spatial and orbital pair structure for the $d_{x^2-y^2}$
gap calculated from the harmonic approximation of $g(k)$ corresponding to $U=1.5$
with a cut-off length of $\lambda_c = 3.3 d_{Fe-Fe}$.}
\label{fig:spatiald}
\end{figure}

 We find that the
off-diagonal amplitudes $a_{nm}$ are negligible compared to the
diagonal ones, and that the orbitals that contribute are the
$d_{xz}$, $d_{yz}$, and $d_{xy}$ orbitals which have weight near
the Fermi surfaces. The amplitudes $a_{nn}(l)$ for the $s$ and the
$d_{x^2-y^2}$ gaps are shown in Figs. \ref{fig:spatials}
and \ref{fig:spatiald}, respectively. %For the $s_\pm$ gap, the
%important diagonal amplitudes involve the $d_{xz}$ and $d_{yz}$
%orbitals, while for the $d_{x^2-y^2}$ gap the $d_{x^2-y^2}$
%orbital contribution is also significant. Figs.~\ref{fig:spatials}
%and \ref{fig:spatiald} show the amplitudes for the $s_\pm$ and
%$d$-wave gaps, respectively.
For the $s$ case, one sees that the internal structure of a pair
consists of a superposition of $(xz\uparrow,xz\downarrow)$
singlets which are formed between a central site and sites
displaced by an odd number of lattice spacings in the
$y$-direction,  $(yz\uparrow,yz\downarrow)$ singlets between the
central site and odd numbered sites in the $x$-direction, and
weaker  $(xy\uparrow,xy\downarrow)$ singlets with a more intricate
$s$-wave arrangement.  The internal structure of the $d_{x^2-y^2}$
state consists of a similar superposition but with a negative
phase difference between the  $(xz\uparrow,xz\downarrow)$ and
$(yz\uparrow,yz\downarrow)$ singlets.  In addition, there is a
significant $d_{x^2-y^2}$ contribution from the
$(xy\uparrow,xy\downarrow)$ singlets confined primarily to the
nearest neighbor sites.

\section{Summary and Conclusions}
\label{sec:8}

We have studied a weak-coupling spin and charge fluctuation model
of the pairing interaction using a tight-binding parametrization of the electronic
band structure obtained from DFT calculations
for  the Fe-pnictides.  We initially reviewed criticism of the
 2-orbital models used in early studies of this kind, and compared
 results with more accurate 5-orbital models.  Within the
 5-orbital framework, we have calculated
 the multiorbital susceptibility.  We calculated
 the static homogeneous spin susceptibility  within this
 framework,
 and showed that it qualitatively agrees with experimental measurements of
 the $T$ dependence, but that an interband susceptibility component
 was necessary to understand the magnitude of the measured low-temperature
 limiting susceptibility.  The $T$-dependence of both $\chi(T)$
 and $(T_1T)^{-1}$ is qualitatively similar to experiments, and
 may be consistent with them for some values of the parameters.  A
 detailed fit was not attempted here.

We then constructed the pairing vertex from the generalized RPA
susceptibilities, and calculated the pairing eigenvalues for
dopings corresponding to undoped and electron-doped materials, for
a variety of interaction parameters corresponding to intra- and
inter-orbital Coulomb matrix elements and Hund's rule couplings.
We found that in the parameter region where there is a significant
coupling strength, the pairing interaction for our model of the
Fe-pnictides arises from the exchange of spin fluctuations.  These
give rise to both intra- and inter-Fermi surface scattering
processes.  In a number of cases the dominant pairing contribution
came from particle-particle scattering processes from the hole
Fermi surface around the $\Gamma$ point to the electron Fermi
surfaces around the $(\pi,0)$ and $(0,\pi)$ points in the unfolded
Brillouin zone. These scatterings involve momentum transfer of
order $(\pi,0)$. We find that there are two leading pairing
channels, one with ``sign-changing" $s$-wave $(A_{1g})$ symmetry
and one with $d_{x^2-y^2}$ $(B_{1g})$ symmetry, and that the gap
functions corresponding to both  have nodes on the Fermi surface.
For values of the interaction parameters corresponding to
significant pairing strength, these two eigenvalues can become
very close. The $s$-wave gap that exhibits a sign change between
its average values on the electron and hole Fermi surfaces
nevertheless also displays nodes on the electron sheet, whereas
the $d$-wave state has its nodes on the hole sheet. The $s$-wave
gap nodes are not required by symmetry, and could be absent for a
different choice of parameters, but appear to be robust within the
manifold of apparently realistic interaction parameters we have
studied.

An obvious question which arises in such studies is the
sensitivity of results to the particular choice of band structure.
We have relied upon the Fermi surface for the paramagnetic system
determined in the Generalized Gradient Approximation (GGA) 
calculations of Cao {\it et al.}~\cite{ref:Cao}, but other band calculations find subtle
differences in Fermi surfaces for the La-1111 material.  We have
attempted to address this question by considering variations of
our effective 5-band model hopping parameters chosen originally to
fit the Cao {\it et al.} band structure. In particular, we
considered variations which led to a perfect nesting of the outer
$\alpha$ and $\beta$ Fermi surface sheets, to try to maximize the
$(\pi,0)$ contribution to the pairing vertex.  We found that our
results for the structure of the susceptibility and pairing
vertices changed in fact very little.  On the other hand, the
subtle changes led to a nearly exact degeneracy of the
extended-$s$ and $d_{x^2-y^2}$ eigenvalues of the linearized
pairing problem.  We argued that this degeneracy becomes exact for
a situation where all sheets have the same radius, and propose
this as a simple explanation of our finding that these two pairing
channels appear to compete very strongly with one another, even in
the realistic cases.

Our results appear to be qualitatively different from
several works\cite{ref:Mazin,ref:Yanagi,ref:Ikeda,ref:Yu,ref:Bernevig}
which find, on heuristic grounds or within microscopic
calculations, that the $A_{1g}$ state is closer to the form $\cos
k_x \times \cos k_y$ (in the unfolded zone), in contrast to our
result, which is closer to the form $\cos k_x + \cos k_y$ in the
sense described above. Although both states have
identical $A_{1g}$ symmetry, the difference between them is
important given the structure of the Fermi surface; for example,
the $\cos k_x \times \cos k_y$ form does not have nodes on either of
the $\beta$ sheets shown in Fig.~\ref{fig:FS2band}. We
believe that the strong pair scattering between the $\beta_1$ and
$\beta_2$ sheets found in our calculations, and neglected in some
simpler models, is crucial to stabilize the gap structure similar
to $\cos kx +\cos ky$.

We further studied the spatial structure of the states
corresponding to the leading pairing eigenvalues. In each case,
the internal structure consists of a superposition of singlets
made up from electrons which occupy the same $d_{xz}$, $d_{yz}$,
or $d_{xy}$ orbitals on different sites.  The distribution and
relative signs of these superpositions are illustrated in
Figs.~\ref{fig:spatials} and \ref{fig:spatiald}.  The fact that the
$s$-wave and $d$-wave solutions have very similar coupling
constants opens the possibility that different members of the
Fe-pnictide superconducting family may have different gap
symmetries.  Furthermore, it suggests that these systems may have
low-lying collective particle-particle modes which would be
$s$-wave like in a $d$-wave superconductor, and $d$-wave like in
an s-wave superconductor~\cite{ref:bardasis}.

There have been several earlier weak-coupling calculations of
pairing in the Fe-pnictide superconductors.  An early suggestion
of an extended $s$-wave sign reversed gap made by Mazin
{\it et al.}~\cite{ref:Mazin} was based upon the Fermi surface structure
found in a DFT calculation.  There it was proposed that $(\pi,0)$
antiferromagnetic spin fluctuations could lead to a
superconducting state with an isotropic $s$-wave gap which had
opposite signs on the electron and hole pockets.

Using a 5-orbital parameterization of the DFT bandstructure,
Kuroki {\it et al.}~\cite{ref:Kuroki} have carried out a
calculation in which an RPA form for the spin and orbital
contributions to the interaction was used along with bare single
particle Green's functions to construct a linearized gap equation.
In a parameter range similar to ours, they found that the leading
pairing instability had $s$-wave symmetry with nodes on the
electron-like Fermi surface.  They also noted that the next
leading channel had $d_{x^2-y^2}$ symmetry and discussed
conditions under which this could become the leading pairing
channel.  The behavior of their $s$-wave gap differs from what we
find by a phase factor of -1 on the electron Fermi surfaces.  That
is, if the gap on the inner hole Fermi surface around the $\Gamma$
point is taken as positive, then Kuroki {\it et al.} find that the
gap on the electron Fermi surface at $(\pi,0)$ reaches its largest
negative value at the point closest to the origin $\Gamma$.  With
the same convention, we find that the gap on the electron Fermi
surface takes on its largest positive value at this point.  This
difference may reflect a difference in the nesting of the hole and
electron Fermi surfaces due to  fits to slightly different band
structures or from the choice of interaction
parameters~\cite{CommentKuroki}.

Wang {\it et al.}~\cite{ref:Wang} have studied the pairing problem using a
5-orbital effective band structure together with the functional
renormalization group approach.  Using the same tight-binding
parameterization as Kuroki {\it et al.}, they also find that the leading
pairing channel has $s$-wave symmetry and that the next leading
channel has $d_{x^2-y^2}$ symmetry.  For their interaction
parameters, they find that there are no nodes on the Fermi
surface, but there is a significant variation in the magnitude of
the gap.  With the sign convention where the gap on the inner hole
Fermi surface is positive, their gap function reaches its smallest
negative value at the point on the electron Fermi surface which is
closest to $\Gamma$.  They note that for other parameter choices,
this $s$-wave gap function on the electron Fermi surface may have
nodes.  This would lead to a gap which becomes positive on a
region of the electron Fermi surface closest to the $\Gamma$
point, in agreement with what we have found.

Both of these studies also conclude that the pairing interaction
arises from spin fluctuation scattering with the dominant
contribution associated with ${\bf Q}\sim (\pi,0)$ scattering of a
pair from the inner Fermi hole surface around the $\Gamma$ point
to the electron Fermi surface around $(\pi,0)$ and $(0,\pi)$. They
also find that the magnitude of the gap on the central hole Fermi
surface is smaller than the maximum magnitudes of the gap on the
inner hole and the electron Fermi surfaces.  Thus the five-orbital 
weak coupling calculations all find singlet pairing in
the extended $s$-wave channel with a nearby $d_{x^2-y^2}$ wave
state. The question of whether there are gap nodes for the
$s$-wave, and indeed whether or not the $s$-wave state is
ultimately stable with respect to the $d$-wave state, would both
appear to depend on parameters.   This sensitivity appears to us
to be a natural consequence of the importance of several orbitals
near the Fermi surface.  In contrast to the cuprates, where
calculations of this kind show a clear $d_{x^2-y^2}$ state well
separated from other pairing states, it seems possible here that
variations in band structure or interaction parameters found in
different materials might possibly lead to different symmetry
ground states.

We close by pointing out that at this writing, experiments have
not conclusively answered either the question of order parameter
symmetry or even whether gap nodes are present.  Early indications
from specific heat~\cite{ref:Mu}, Andreev point contact
spectroscopy~\cite{ref:Shan}, and NMR~\cite{ref:Ning} on the 1111
materials suggested nodes in the superconducting order parameter.
On the other hand, some penetration depth
experiments~\cite{ref:Hashimoto} and ARPES
experiments~\cite{ref:Ding,ref:Evtushinsky} appear to find nearly
isotropic gaps in the 122 materials.  Our calculations find nearly
degenerate $s$ and $d_{x^2-y^2}$ states at $T_c$, both of which
have nodes on the Fermi surface. If the  experiments indicating a
lack of nodes are correct, there are two ways to imagine
reconciling our conclusions with experiments on the 122 materials.
The first is to consider the effects of disorder present at fairly
high levels in current crystals, which should average the order
parameters to a finite quasiisotropic value in the $s$ case. The
second possibility follows from the observation that if the
$d_{x^2-y^2}$ state is the leading eigenvalue at $T_c$, the
thermodynamic ground state may be unstable towards an admixture of
the $s$ state\cite{ref:sczhang_spid}.  A similar phenomenon occurs
in the phase diagram of ordinary $s$ and $d$ symmetries in
one-band superconductors with these pairing channels
competing~\cite{ref:joynt}. Separating these possibilities may be
difficult, and will probably require phase sensitive probes as
were discussed in the cuprate context~\cite{ref:tsuei}.

\begin{acknowledgments}
This work is supported by DOE DE-FG02-05ER46236 (PJH).
SG gratefully acknowledges support by the
Deutsche Forschungsgemeinschaft. DJS and TAM acknowledge the
Center for Nanophase Materials Sciences, which is sponsored at
Oak Ridge National Laboratory by the Division of Scientific
User Facilities, U.S. Department of Energy.
We acknowledge useful discussions with 
O.K.~Andersen, A.~Chubukov, T.~Imai, I.~Mazin, X.-L.~Qi and S.~Raghu.
We particular thank S.~Raghu for his insight regarding the interaction matrices.
\end{acknowledgments}

\vspace{2em}
\appendix*
\section{Model parameters for the 5 band fit}
\label{appendix}

In the following, the model parameters for the 5 band tight
binding fit of the DFT band structure by Cao {\it et al.} are
listed. Including hopping up to fifth nearest neighbors on an
effective Fe-Fe lattice we find intraorbital kinetic energy terms
$\xi_{mm} (k)$ and interorbital kinetic terms $\xi_{mn} (k)$ that
respect the basic symmetry requirements imposed by the point group
of the crystal. The intra- and interband hopping parameters are
given in Tab.~\ref{tab:intraorbitalhop} and
Tab.~\ref{tab:interorbitalhop}, respectively. The onsite energies
$\epsilon_i$ (in eV) $\epsilon_1= 0.13$, $\epsilon_3=
-0.22$, $\epsilon_4= 0.3$, and $\epsilon_5= -0.211$ where $i=1$
corresponds to the $d_{xz}$, $i=2$  to the $d_{yz}$, $i=3$ to the
$d_{x^2-y^2}$, $i=4$ to the $d_{xy}$, and $i=5$  to the
$d_{3z^2-r^2}$ orbital, and the band structure is listed below.
For the toy model with nearly spherical Fermi surface sheets we
have changed $t_{xy}^{11}$ from $0.28$ to $0.3$ and $t_x^{13}$
from $-0.354$ to $-0.409$. This is sufficient to produce nearly
degenerate and spherical $\alpha_2$ and $\beta$ FS sheets.
For the Coulomb interaction strengths we have used,
the bare ratios $U/t_{xy}^{11}$ and $U/t_{x}^{13}$ are of order 5,
comparable to values of $U/t$ in the cuprates.  However, we
believe that for the Fe-pnictide system, the interaction is
"spread out" across multiple bands leading to a subtantially less
strongly correlated system\cite{ref:anisimov}.

\begin{widetext}
\begin{eqnarray}
\xi_{11/22}  & = & 2 t_{x/y}^{11} \cos k_x + 2 t_{y/x}^{11} \cos k_y + 4 t_{xy}^{11} \cos k_x \cos k_y
\pm 2 t_{xx}^{11} (\cos 2 k_x -\cos 2 k_y) + 4 t_{xxy/xyy}^{11} \cos 2 k_x \cos k_y  \nonumber \\
& & + 4 t_{xyy/xxy}^{11} \cos 2 k_y \cos k_x + 4 t_{xxyy}^{11} \cos (2 k_x) \cos (2 k_y) \nonumber \\
\xi_{33}  & = & 2 t_{x}^{33} (\cos k_x + \cos k_y) + 4 t_{xy}^{33} \cos k_x \cos k_y
+ 2 t_{xx}^{33} (\cos 2 k_x + \cos 2 k_y)  \nonumber \\
\xi_{44}  & = & 2 t_{x}^{44} (\cos k_x + \cos k_y) + 4 t_{xy}^{44} \cos k_x \cos k_y
+ 2 t_{xx}^{44} (\cos 2 k_x + \cos 2 k_y)  \nonumber \\
& & + 4 t_{xxy}^{44} (\cos 2 k_x \cos k_y + \cos 2 k_y \cos k_x)  + 4 t_{xxyy}^{44} \cos 2 k_x \cos 2 k_y  \nonumber \\
\xi_{55}  & = & 2 t_{x}^{55} (\cos k_x + \cos k_y) + 2 t_{xx}^{55} (\cos 2 k_x + \cos 2 k_y) \nonumber \\
& & + 4 t_{xxy}^{55} (\cos 2 k_x \cos k_y + \cos 2 k_y \cos k_x)  + 4 t_{xxyy}^{55} \cos 2 k_x \cos 2 k_y \nonumber \\
\xi_{12}  & = &  - 4 t_{xy}^{12} \sin k_x \sin k_y - 4 t_{xxy}^{12} (\sin 2 k_x \sin k_y + \sin 2 k_y \sin k_x)
- 4 t_{xxyy}^{12} \sin 2 k_x \sin 2 k_y  \nonumber \\
\xi_{13/23}  & = &  \pm 2 i t_x^{13} \sin k_{y/x} \pm 4 i t_{xy}^{13} \sin k_{y/x} \cos k_{x/y}
    \mp 4i t_{xxy}^{13} (\sin 2 k_{y/x} \cos k_{x/y} - \cos 2 k_{x/y} \sin k_{y/x}) \nonumber \\
\xi_{14/24}  & = & 2 i t_x^{14} \sin k_{x/y} + 4 i t_{xy}^{14} \cos k_{y/x} \sin k_{x/y} + 4i t_{xxy}^{14}
\sin 2 k_{x/y} \cos k_{y/x} \nonumber \\
\xi_{15/25}  & = & 2 i t_x^{15} \sin k_{y/x} - 4i t_{xy}^{15} \sin k_{y/x} \cos k_{x/y}  - 4i t_{xxyy}^{15}
\sin 2 k_{y/x} \cos 2 k_{x/y}  \nonumber \\
\xi_{34}  & = & 4 t_{xxy}^{34} (\sin 2 k_y \sin k_x  - \sin 2 k_x \sin k_y ) \nonumber \\
\xi_{35}  & = & 2 t_x^{35} (\cos k_x - \cos k_y) + 4 t_{xxy}^{35} (\cos 2 k_x \cos k_y - \cos 2 k_y \cos k_x) \nonumber \\
\xi_{45}  & = & 4 t_{xy}^{45} \sin k_x \sin k_y + 4 t_{xxyy}^{45} \sin 2 k_x \sin 2 k_y \nonumber
\end{eqnarray}
\end{widetext}

\begin{table}
\begin{tabular}{|c|c|c|c|c|c|c|c|}
\hline
$t_{i}^{mm}$ & $i=x$ & $i=y$ & $i=xy$ & $i=xx$ & $i=xxy$ & $i=xyy$ &$i=xxyy$ \tabularnewline
\hline
$m=1$ & $-0.14$ & $-0.4$ & $0.28$ & $0.02$  & $-0.035$ & $0.005$ & $0.035$ \tabularnewline
\hline
$m=3$ & $0.35$ & & $-0.105$ & $-0.02$ & & & \tabularnewline
\hline
$m=4$ & $0.23$ & & $0.15$ & $-0.03$  & $-0.03$ & & $-0.03$ \tabularnewline
\hline
$m=5$ & $-0.1$ & & & $-0.04$  & $0.02$ & & $-0.01$ \tabularnewline
\hline
\end{tabular}
\caption{The intraorbital hopping parameters used for the DFT fit of the 5 orbital model.
\label{tab:intraorbitalhop}}
\end{table}

\begin{table}
\begin{tabular}{|c|c|c|c|c|}
\hline
$t_{i}^{mn}$ & $i=x$ & $i=xy$ & $i=xxy$ & $i=xxyy$ \tabularnewline
\hline
$mn=12$ & & $0.05$ & $-0.015$ & $0.035$ \tabularnewline
\hline
$mn=13$ & $-0.354$ & $0.099$ & $0.021$ & \tabularnewline
\hline
$mn=14$ & $0.339$ & $0.014$ & $0.028$  & \tabularnewline
\hline
$mn=15$ & $-0.198$ & $-0.085$ & & $-0.014$ \tabularnewline
\hline
$mn=34$ &  &  & $-0.01$ & \tabularnewline
\hline
$mn=35$ & $-0.3$ &  & $-0.02$  & \tabularnewline
\hline
$mn=45$&  & $-0.15$ & & $0.01$ \tabularnewline
\hline
\end{tabular}
\caption{The interorbital hopping parameters used for the DFT fit of the 5 orbital model.
\label{tab:interorbitalhop}}
\end{table}

\end{document}